\documentclass[fleqn,usenatbib]{mnras}

\usepackage{newtxtext,newtxmath}

\usepackage[T1]{fontenc}
\usepackage{tikz}
\usetikzlibrary{positioning, arrows.meta}

\DeclareRobustCommand{\VAN}[3]{#2}
\let\VANthebibliography\thebibliography
\def\thebibliography{\DeclareRobustCommand{\VAN}[3]{##3}\VANthebibliography}

\usepackage{graphicx}	
\usepackage{amsmath}	
\usepackage{hyperref}
\usepackage{orcidlink}

\definecolor{bg_cnn}{HTML}{99DDFF} 
\definecolor{bg_drop}{HTML}{FFAABB} 
\definecolor{bg_fc}{HTML}{77AADD} 
\definecolor{bg_concat}{HTML}{44BB99} 
\definecolor{bg_io}{HTML}{DDDDDD} 

\title[GALAH DR4 europium abundances]{A value-added catalogue of neural network-based europium abundances for GALAH DR4}

\author[S. G. Kane et al.]{Sarah G. Kane\orcidlink{0000-0001-8411-1012}$^{*1}$\thanks{E-mail: sgk27@cam.ac.uk (SGK)}, Zofia Kaczmarek\orcidlink{0009-0007-4089-5012}$^{*2}$\thanks{E-mail:  zofia.kaczmarek@uni-heidelberg.de (ZK)}, Andrew Garner\orcidlink{0009-0004-2599-2293}$^{3}$, Sven Buder\orcidlink{0000-0002-4031-8553}$^{4}$, Stephanie Monty\orcidlink{0000-0002-9225-5822}$^{5,6}$,\newauthor and Elana Kane\orcidlink{0009-0009-1988-7047}$^1$\\
$^{1}$Institute of Astronomy, University of Cambridge, Madingley Rd, Cambridge CB3 0HA, UK\\
$^{2}$Zentrum f{\"u}r Astronomie der Universit{\"a}t Heidelberg, Astronomisches Rechen-Institut, M{\"o}nchhofstr. 12-14, 69120 Heidelberg, Germany\\
$^{3}$School of Mathematics \& Physics, University of Surrey, Guildford GU2 7XH, UK\\
$^{4}$Research School of Astronomy and Astrophysics, Australian National University, Canberra, ACT 2611, Australia\\
$^{5}$Center for Interdisciplinary Exploration and Research in Astrophysics (CIERA), Northwestern University, 1800 Sherman Avenue, Evanston, IL 60201, USA \\
$^{6}$New Mexico State University, Department of Astronomy, 1320 Frenger Mall, Las Cruces, NM 88003-8001, USA}

\date{Accepted XXX. Received YYY; in original form ZZZ}

\pubyear{\the\year{}}

\begin{document}
\label{firstpage}
\pagerange{\pageref{firstpage}--\pageref{lastpage}}
\maketitle

\begin{abstract}
The rapid neutron-capture ($r$-process) element europium (Eu) is a valuable tracer of neutron star mergers and other rare nucleosynthetic events. The stellar spectroscopic survey GALAH's unique wavelength range and setup include the Eu absorption feature at $\sim6645$~\AA~for almost a million stars in the most recent Data Release 4 (DR4). However, DR4 also saw a decreased precision in reported Eu measurements compared to previous data releases. In this work, we use a convolutional neural network (CNN) to perform label transfer, wherein we use the GALAH DR4 spectra and stellar parameters to infer DR3 [Eu/H] abundances. This CNN is then applied to DR4 spectra without corresponding DR3 Eu abundances to develop a new, publicly available catalogue of [Eu/H] values for high signal-to-noise targets. We include [Eu/H] predictions for $118\,946$ stars, out of which $54\,068$ giants constitute our ``golden sample'' of high-confidence predictions, which pass stricter quality cuts and have a reported precision $\lesssim0.1$. To overcome the scarcity of training data in the low metallicity regime, we provide an additional catalogue of [Eu/H] abundances for metal poor ($\mathrm{[Fe/H]}<-1$) stars derived from synthesis of the Eu feature. Our ``golden sample’' can be combined with [Eu/H] values from GALAH DR3 to create a catalogue of over $100\,000$ vetted, high-quality abundances on a homogeneous scale. Moreover, we are able to reproduce known science results, including the elevated Eu abundances of accreted stars and previously observed Galactic chemical evolution trends. This catalogue represents one of the largest available samples of [Eu/H] abundances for high signal-to-noise targets.
\end{abstract}



             
\def\thefootnote{*}\footnotetext{These authors contributed equally to this work and share joint first authorship.}\def\thefootnote{\arabic{footnote}}

\section{Introduction}

Ushered in by the first discovery of gravitational waves \citep{Abbott_2016_GW_discovery}, we have now entered a new era of multi-messenger astronomy. Since 2015, hundreds of black hole-black hole (BH) and BH-neutron star (NS) mergers have been observed via their gravitational wave signatures \citep[see e.g.,][]{LIGO_2023_summary}, yet only two binary neutron star mergers (NSMs) have been found thus far \citep{GW170817_NSM_discovery, GW190425_NSM_discovery_2020}. With only a scarcity of data from gravitational waves available for these events, elements formed via the rapid-neutron capture process (\textit{r}-process) promise to be a key piece of evidence as a complimentary channel to gravitational wave events, as they are formed in neutron star mergers. In fact, \textit{r}-process production has \emph{only} been directly observed occurring in a kilonova, the electromagnetic counterpart to a NSM \citep{Abbott_2017_NSM, Smartt_2017_rProcess,Drout_2017}.

However, although the connection between NSMs and the \textit{r}-process has been confirmed, chemical evolution models have yet to unanimously constrain the sites and rates of \textit{r}-process nucleosynthesis to match observations in the Milky Way (MW). Europium, a second peak \textit{r}-process element, is among the easier of such elements to measure because of the existence of a redder line (which is used in this work), and thus many studies look at the Eu content of stars as a tracer of the \textit{r}-process \citep[e.g.,][]{Simonetti_2019,Matsuno_2021,Monty_2024}. With stars forming from the material enriched by previous generations of nucleosynthetic events, we can thus begin to reconstruct the \textit{r}-process production history of the MW via the europium we observe in stars today.
A distinct pattern in [Eu/Fe] abundances has been observed relative to stellar metallicity, [Fe/H], which correlates with age in systems with a single star formation history. In particular, at low metallicities ([Fe/H]~$\lesssim-1$), there is a substantial scatter in stellar [Eu/Fe], reflecting the stochasticity of \textit{r}-process events \citep{Haynes_Kobayashi_2019,vandeVoort2020}. It is in these metal-poor halo populations that the vast majority of Eu-enhanced stars ([Eu/Fe]~$>0.3$) have been observed \citep[see][and references therein]{Frebel_review_2023}. By comparison, there is a ``knee'' in the observed stellar [Eu/Fe] at [Fe/H]~$\approx-1$, with [Eu/Fe] values steadily decreasing with increasing metallicity beyond this point \citep[e.g.,][]{Battistini_Bensby_2016}.
The Galactic disk, which formed at [Fe/H]~$\approx-1$ \citep{Belokurov2022, Chandra_2024, Zhang_2024, Conroy_2022}, is thus uniformly low in europium, with only a rare few Eu-rich stars discovered \citep[e.g.,][]{Xie_2024,Xie_2025}. Because there is a significant inherent delay time in the formation of a neutron star binary and their subsequent merging, chemical enrichment models typically must invoke an additional prompt enrichment source \citep{Wehmeyer_2015,Simonetti_2019,Cote_2019,Skuladottir_2020,Molero_2023,Chen_2025, Henderson_2025}, often rare magneto-rotational core collapse supernovae, or metallicity-dependent rates for NSMs \citep[e.g.,][]{Simonetti_2019,Kobayashi_2023} to match the Eu abundance patterns observed in stars at low metallicities.

Although Eu remains one of the \emph{easier} \textit{r}-process elements to observe, abundance determinations remain challenging. Many large stellar spectroscopic surveys, such as APOGEE \citep{APOGEE_survey,APOGEE_DR17}, DESI \citep{DESI_MW, DESI_EDR_MW}, or LAMOST \citep{LAMOST, LAMOST_LEGUE}, lack either the wavelength range, the resolution, or both to measure Eu lines. Notably, the Gaia-ESO survey \citep{Gaia_eso_survey, Gaia_ESO_results} \emph{does} include Eu abundances, but it observes only $114\,916$ stars, not all of which will have robust measurements. Because of the difficulties associated with determining Eu and the inherent utility of this element as a tracer of rare nucelosynthetic events, the community has also relied on targeted studies of Eu abundances, especially in the halo, such as the \textit{r}-Process Alliance \citep[RPA, e.g.,][]{RPA_DR1, RPA_Sakari_2018, RPA_DR4_2020, RPA_DR5_2024}.
Among large stellar spectroscopic surveys, GALactic Archeology with HERMES \citep[GALAH,][]{GALAH_scientific_motivation} remains unique as its extent into blue optical wavelengths and spectral resolution of $R\approx28\,000$ allows for neutron-capture abundance determination, including the Eu abundance via the line at $6645.11$~\AA. The most recent Data Release 4 \citep[DR4,][]{GALAH_DR4} of GALAH includes $917\,588$ stars, and even if one restricts this sample to targets with the high-SNR observations (e.g., SNR $> 50$) often necessary for robust Eu line measurements, more than $100\,000$ stars remain.

\begin{figure}
    \includegraphics[width=0.85\columnwidth, alt={In the representation of GALAH DR3 versus DR4 [Eu/H] values, the abundances determined by the two data releases appear essentially entirely uncorrelated and appear as a "cloud" of unrelated values around the 1-to-1 line. Most [Eu/H] values fall between -0.5 and 1, and there is a sharp cutoff in [Eu/H] abundances at 1 for GALAH DR4. The R-squared value listed in the figure is -2.10.}]{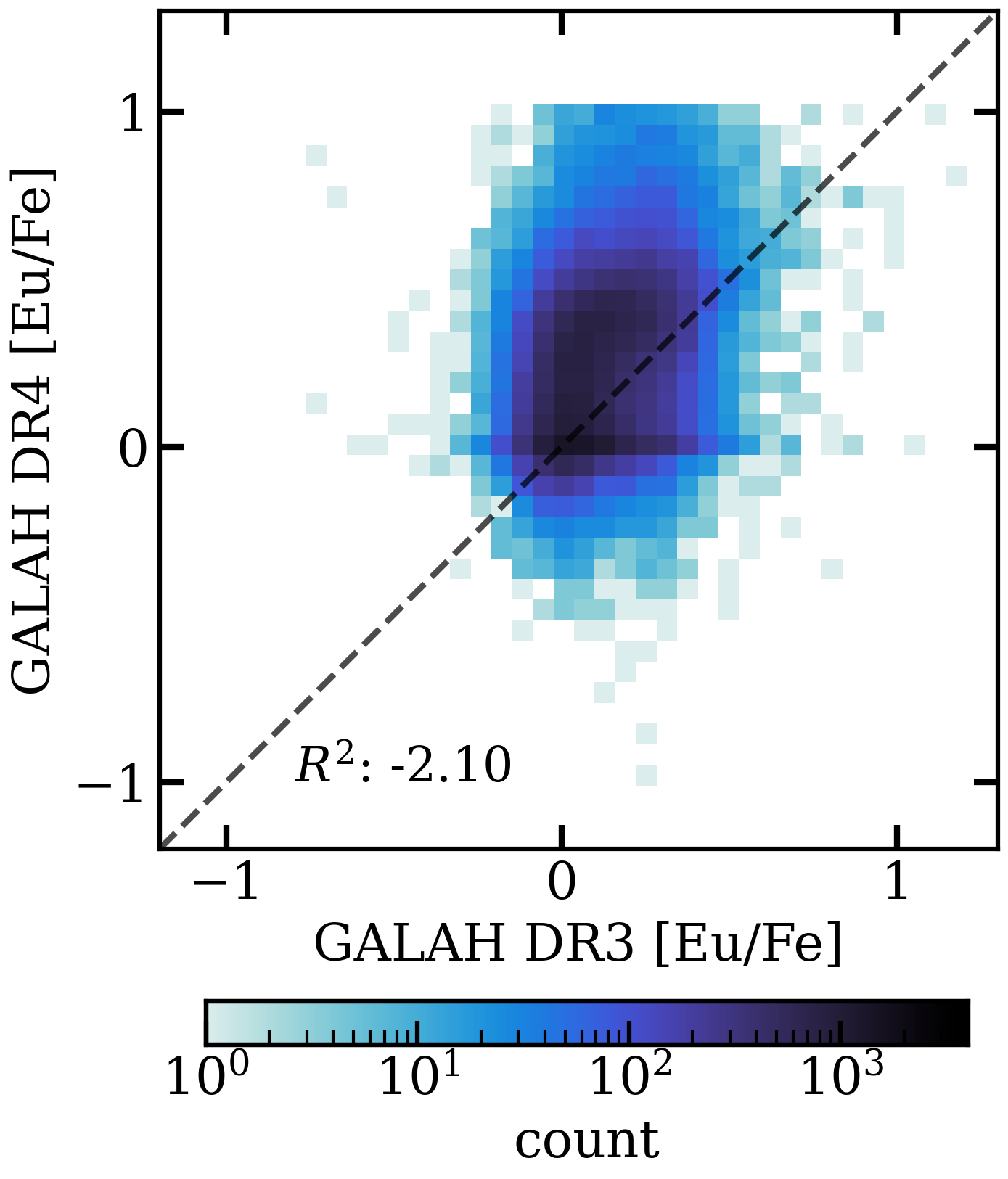}
    \caption{The GALAH DR4 versus GALAH DR3 [Eu/Fe] measurements for a subsample of stars with GALAH best practice cuts (\url{https://www.galah-survey.org/dr3/using_the_data/}), in log-scaled density of the number of stars. The black dashed diagonal line marks the 1:1 relation for illustrative purposes; the sharp cut at [Eu/Fe] = 1 is a feature of the GALAH DR4 pipeline. The $R^2$ score is listed in the lower left corner, demonstrating that GALAH DR4 [Eu/Fe] is not a good predictor of GALAH DR3 [Eu/Fe].}
    \label{fig:DR3_DR4_Eu}
\end{figure}

Although previous data releases of GALAH had Eu abundances that were broadly community-tested and used \citep[see][among many others which used GALAH Eu measurements]{Matsuno_2021,Aguado_2021,Buder_2022, Monty_2024,Manea_2024}, DR4 saw a decreased accuracy in Eu measurements. This fact is illustrated in Fig.~\ref{fig:DR3_DR4_Eu}, which shows GALAH DR4 versus DR3 reported values for [Eu/Fe] for a sample of stars in both releases. As is clear in the figure, there is no consistency between the DR3 and DR4 abundances\footnote{In the context of machine learning, an $R^2$ value describes how well the predicted label is explained by the model. $R^2=1$ thus indicates perfect model predictions, and $R^2=0$ would be the value obtained by a model that ``guesses'' the average value of the label for every prediction. $R^2<0$ is thus possible for models which make predictions worse than the average label.}, and the GALAH DR4 release paper also reports a decreased precision in Eu abundances \citep{GALAH_DR4}. That differences arose between GALAH DR3 and DR4 is likely caused by the substantial change in the methodology between the releases.
In particular, while GALAH DR3 \citep{GALAH_DR3} derived abundances from spectral synthesis with Spectroscopy Made Easy \citep[SME,][]{Piskunov2017}, GALAH DR4's abundances \citep{GALAH_DR4} were determined via a neural network trained on synthetic spectra. Europium may have been especially affected by these changes because there is essentially only one usable line in the GALAH wavelength range (at $6645.11$~\AA), which is not exceptionally strong, blended with a weak Fe line, and impacted by isotopic splitting and hyperfine structure effects. These factors likely compounded to result in the difficulties with [Eu/Fe] determination across the vast parameter space of GALAH DR4.

The crux of this work hinges on the fact that GALAH DR3's Eu abundances were community-tested and trusted, and so for the remainder of this work we take them to be the ``ground truth'' Eu values. We are thus motivated by a goal to move GALAH DR4 [Eu/H] abundances onto a ``DR3 scale''. That is, we want to perform a label transfer of DR3 [Eu/H] abundances onto spectra that are exclusive to GALAH DR4, thus constructing a new catalogue of europium abundances for the data release that is benchmarked to GALAH DR3. We choose to use [Eu/H] as the label for transfer rather than [Eu/Fe] due to differences between GALAH DR3 and DR4 [Fe/H], which diverge by an average $0.0855$~dex.
At [Fe/H]~$\lesssim-0.8$, these differences become an increasingly systematic offset of $\sim0.14$~dex. Using [Eu/H] as our label thus allows us to accommodate differences in metallicities between the data releases without skewing our results.

To infer GALAH DR3-scale europium abundances for GALAH DR4 spectra, we use a convolutional neural network (CNN) constructed with {\tt PyTorch} \citep{Paszke_pytorch}. CNNs have previously been used to infer abundances from spectra, showing promising efficiency and precision compared to classical methods \citep[e.g.,][]{Wang2020, RAVE_CNN, Guiglion2024, Ambrosch2023, Liu2025}. Our work here fits within a broader and ever-growing range of literature on data-driven abundances, including tools such as the Cannon \citep{Ness_cannon}, the Payne \citep{Payne_YSTing_2019}, and astroNN \citep{AstroNN_Leung_Bovy_2019}. 

This work is organized as follows. In Section~\ref{subsec:data}, we describe our data: the training/validation dataset from a GALAH DR3/DR4 crossmatch and the sample of GALAH DR4 stars without DR3 labels. In Section~\ref{subsec:methods}, we outline the procedure by which we perform the label transfer of GALAH DR3 [Eu/H] abundances to GALAH DR4 spectra using a CNN. In Section~\ref{sec:validation}, we present our results. Sections~\ref{subsec:validation_performance}-\ref{subsec:korg} contain a thorough discussion of network performance on validation datasets, including feature importance analysis. 
Because metal-poor stars (with [Fe/H]~$<-1$) are of broad community interest and are simultaneously more challenging for the CNN to produce accurate predictions, we give them a more detailed discussion and also develop a secondary catalogue of abundances from spectral synthesis in Section~\ref{subsec:metal_poor}. We propose quality cuts and best practices in Section~\ref{subsec:data_cuts}, and select a ``golden sample'' of most reliable predictions in Section~\ref{subsec:golden_sample}.
We describe our rationale for using a data-driven approach in Section~\ref{subsec:why_NN}, and include
additional science use-case tests in Section~\ref{subsec:science_validation} to provide further verification. Our conclusions are given in Section~\ref{sec:conclusions}.

\section{Data}
\label{subsec:data}

\subsection{Training and Validation Data}

\label{subsec:test_train}

\begin{figure}
	\includegraphics[width=\columnwidth, alt={The five example spectra include four prominent absorption features. The strongest is the nickel line at 6643.63 angstroms, which has an approximate depth of 0.4 in normalized flux. Next, there is the important europium feature at 6645.11 angstroms, which has a depth of approximately 0.05 to 0.1. There are two prominent iron lines at 6646.93 and 6648.08 angstroms, which have depths of approximately 0.05 and 0.1, respectively. The spectra all look very similar with the exception of the europium feature, which becomes stronger in the spectra of more Eu-enhanced stars. The [Eu/H] values range from approximately -1.0 to -0.4.}]{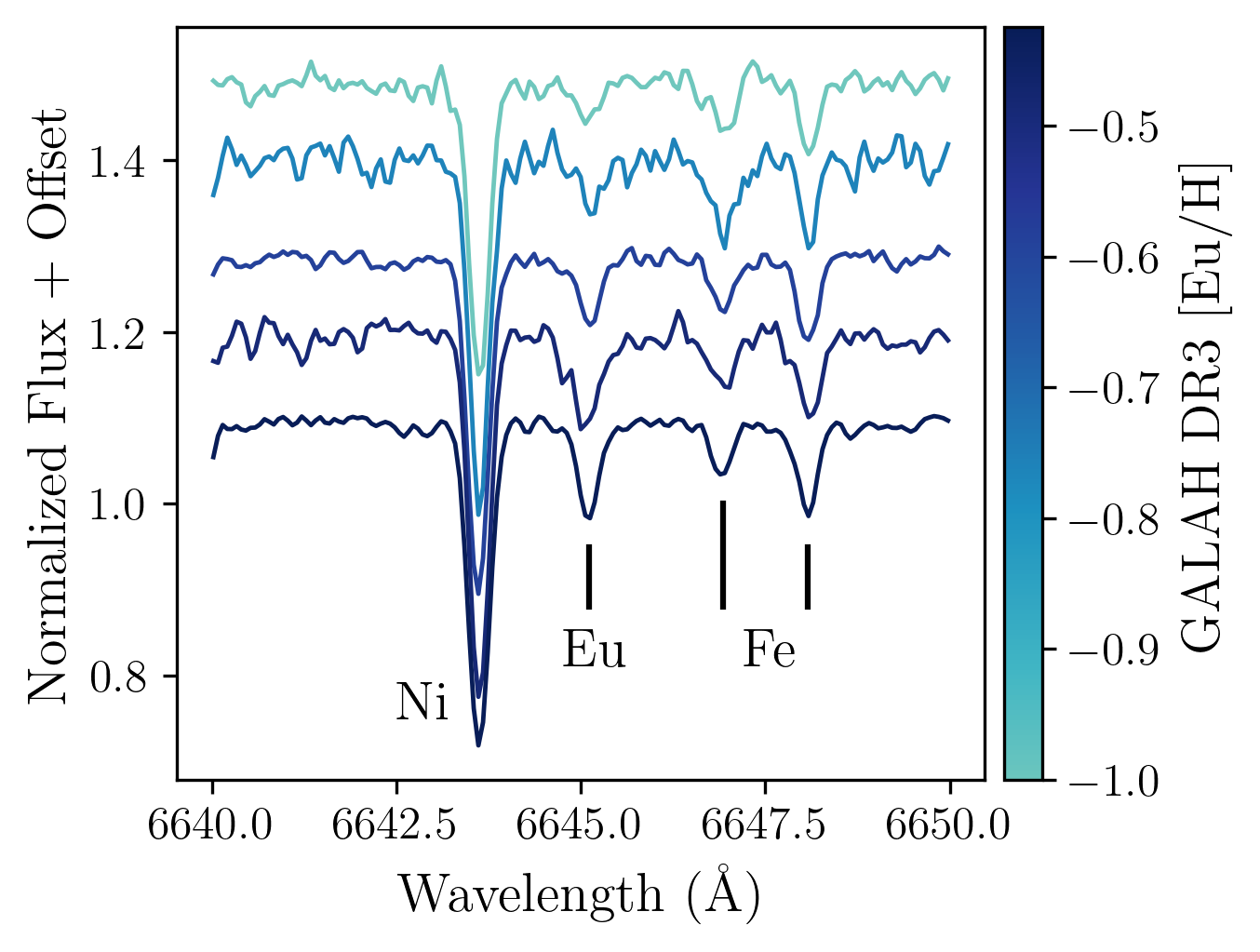}
    \caption{A set of five example spectra from GALAH DR4 in the wavelength range used by the neural network in this work ($\sim6640$~\AA~to $\sim6650$~\AA), with varying [Eu/H] values from GALAH DR3 indicated by the colorbar. A vertical offset of 0.1 is applied to the normalized flux for easier visualization. Each of these stars have similar stellar parameters in GALAH DR3, with $T_\textrm{eff}\approx4450$~K, $\log g\approx1.65$, and [Fe/H]~$\approx-1.1$. The spectra are all interpolated onto our uniform grid of wavelengths. for input with the neural network. The Eu line as well as the Ni and two Fe lines within the spectra are marked.}
    \label{fig:example_spectra}
\end{figure}

\begin{figure*}
    \includegraphics[width=2\columnwidth, alt={The figure shows three 2D histograms for the distribution of our train and test data in the surface gravity versus effective temperature space, with each bin in the histogram color-coded to indicate the number of stars, the mean GALAH DR3 [Fe/H], and the mean GALAH DR3 [Eu/H]. Effective temperatures fall between approximately 4000 and 8000 Kelvin, and surface gravities range between about 0 and 5. The main sequence (dwarf stars) is located at surface gravities greater than 3.5 for stars hotter than 5500 K or  greater than 3.8 for stars cooler than 5500 K; stars below these surface gravities are giants. Most stars in the training and validation data are located on the giant branch, with only a few thousand stars located along the main sequence. The few vert hot main sequence stars are excluded via the cut for temperatures less than 6600 Kelvin. These excluded hot dwarfs all appear to have average [Fe/H] greater than 0 and [Eu/H] greater than 1. Along the main sequence and down to surface gravities among the giants of about 2.5, most stars have [Fe/H] between about -0.8 and -0.2. Typical [Eu/H] values among these stars look to be between approximately -0.5 and 0.5. For giants with surface gravities less than about 2.5, especially the hotter of these stars, [Fe/H] can be on average between -2 and -1, and [Eu/H] averages are between about -0.4 and 0.7. Most metal-poor stars are thus closer to the top of the red giant branch.}]{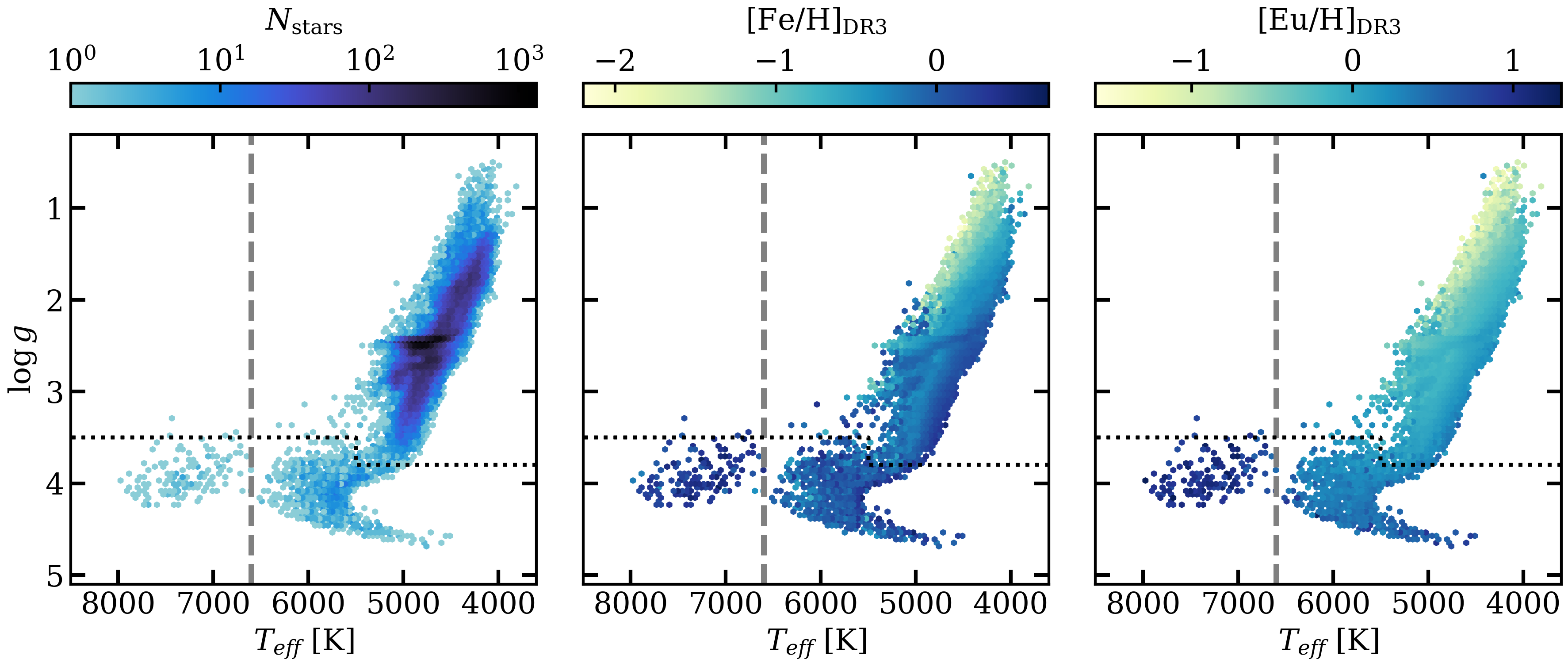}
    \caption{The three subplots illustrate the same sample of $52\,147$ stars used for the training and validation datasets ($52\,302$ including the $T_{\rm eff} > 6600$ K stars seen left of the dashed grey vertical line and 3 stars with corrupted Eu lines, which were eventually removed from the datasets), coloured by density (left) or chemical information (center/right). The dotted black line shows a boundary between GALAH DR3 dwarfs and giants from \citet{Borisov2022} used in the train-test split. \textit{Left:} A 2D hexagonal binning of the stars on the Kiel diagram, coloured by density (the number of stars per bin). \textit{Center:} As left, but coloured by average [Fe/H] within bins. \textit{Right:} As left, but coloured by [Eu/H] values derived from GALAH DR3.}
    \label{fig:train_test}
\end{figure*}

Given that the goal of this work is to perform a label-transfer of GALAH DR3 [Eu/H] values to GALAH DR4 spectra, we require a sample of stars with high-quality DR3 europium abundances upon which to train and validate our model \citep[for more details on the abundance derivation, see GALAH DR3 in][]{GALAH_DR3}. Per the GALAH survey's recommended best practices\footnote{\href{https://www.galah-survey.org/dr3/using_the_data/}{GALAH DR3 best practices} and \href{https://www.galah-survey.org/dr4/using_the_data/}{GALAH DR4 best practices}}, we require that there are no errors flagged in the stellar parameter determination in DR3 or DR4, which is denoted by $\texttt{flag\_sp}=0$ in both. In DR4 we also use only stars with $\texttt{flag\_sp\_fit}=0$ for good stellar parameter fits. We make a cut for good metallicity ([Fe/H]) values with $\texttt{flag\_fe\_h}=0$ only in GALAH DR3 and omit this requirement in DR4, per GALAH recommendations.
We also require that there are no flagged errors in the spectrum reduction pipeline, corresponding to $\texttt{red\_flag}=0$ in the GALAH DR3 error flags and $\texttt{flag\_red}=0$ in DR4.
As the Eu line at $6645.11$~\AA~is not overly strong and following from other works that use GALAH DR3 europium measurements \citep{Matsuno_2021}, we restrict our training data to those with signal-to-noise ratios great than 50 on CCD3, where the Eu line is located ($\texttt{snr\_c3\_iraf}>50$ in DR3 and $\texttt{snr\_px\_ccd3}>50$ in DR4). Furthermore, we require that there are no errors in the DR3 [Eu/Fe] abundance measurements with $\texttt{flag\_Eu\_fe}=0$. Finally, we exclude the hottest stars ($\texttt{teff}>6600$ K), as they systematically show unrealistically high DR3 [Eu/H] values (see Fig.~\ref{fig:train_test}); this has a minimal (-0.3\% of stars) effect on the dataset size.

We use the spectra themselves to infer new $\textrm{[Eu/H]}$ values. These spectra, as published in GALAH DR4, are already continuum normalized and radial velocity-corrected. However, the radial velocity correction necessarily shifts the wavelength values for each pixel, so to give our neural network a consistent set of data, we interpolate the full CCD3 spectra onto a uniform of grid of 4096 wavelength values within the $6481-6739$~\AA~extent.

Given that we \emph{only} seek to predict Eu abundances, we select a $10$~\AA~region of the spectrum to use ranging from $6640$~\AA~to $6650$~\AA~(corresponding to 159 gridpoints). This wavelength window includes the necessary europium line at $6645.11$~\AA~as well as a nickel line and two prominent iron lines; a weaker iron line at $6645.37$~\AA~is also blended with the Eu feature \citep{Arcturus_atlas}. To model this region, we thus include Eu, Ni, and Fe abundances as model parameters. Our choice to use only a narrow wavelength region centered on the Eu line is well-motivated by the fact that \citet{GALAH_DR4} hypothesize that the GALAH DR4 Eu measurements may have decreased accuracy because the full-spectrum fitting used in their methodology failed to adequately capture the single Eu feature. Several sample spectra are illustrated in Fig.~\ref{fig:example_spectra} for stars of varying europium abundances, with the absorption features noted.

Visual inspection of the spectra revealed that a number of them contained pixels with anomalously high or low normalized fluxes. We found $5\sigma$-clipping and replacing the corrupted gridpoints with the $\lambda$-wise median value worked well for alleviating this issue. However, further inspection of the modified spectra revealed this treatment also overwrote some rare intrinsic features, typically for spectra containing $5-10\sigma$ outliers or mismatched flux normalization. To target primarily extrinsic artefacts (e.g. cosmic rays), we apply clipping only to spectra containing any $>10\sigma$ outliers and with flux continuum level within $\pm0.07$ of 1.0. We also entirely reject three spectra where the Eu line itself is strongly corrupted (i.e. any value within $\Delta\lambda = 0.15$~\AA~of $6645.11$~\AA~would be overwritten) from the training/validation dataset. Following all cuts, our dataset contains $52\,147$ stars, which is then split randomly into an $80\%$ training set for the neural network with $20\%$ withheld for validation testing. To help with accurate representation of different stellar types in the training dataset, we ensure both the dwarf and giant subsets (divided by the dotted line in Fig.~\ref{fig:train_test}) are split $80\%-20\%$ into the training and validation sets.

In summary: the input features for our neural network are the RV-corrected, grid-interpolated spectra released from GALAH DR4 in the wavelength range of $6640$~\AA~to $6650$~\AA~as well as the DR4 reported stellar parameters: $T_\textrm{eff}$, $\log g$, $\textrm{[Fe/H]}$ and [Ni/Fe]. The choice of stellar parameters was motivated by a feature importance analysis, which we discuss in Section~\ref{sec:feature_importance}. The label predicted by the network is then the europium abundance relative to hydrogen, $\textrm{[Eu/H]}$, on the Solar scale $\mathrm{A(Eu)} = \log_{10}(N_\mathrm{Eu}/N_\mathrm{H}) + 12= 0.57$ \citep{GALAH_DR3}. For the training and validation data, $\textrm{[Eu/H]}$ is derived from GALAH DR3 catalogue's Fe and Eu abundances: $\textrm{[Eu/H]}=\textrm{[Eu/Fe]}_\textrm{DR3}+\textrm{[Fe/H]}_\textrm{DR3}$.

To avoid irregular network behaviour, all inputs (interpolated spectra and stellar parameters) and labels are rescaled to unit variance using the {\tt StandardScaler} implemented in {\tt scikit-learn} \citep{scikit-learn}. We visualize the training and validation datasets on a Kiel diagram of effective temperature and surface gravity in Fig.~\ref{fig:train_test}.

\subsection{GALAH DR4 without DR3 Labels}
\label{subsec:prediction_set}
The GALAH DR3-DR4 cross-match selection described in Section~\ref{subsec:test_train} serves as our training and validation data, but the ultimate goal of this work is to infer reliable Eu abundances from GALAH spectra that have \emph{no} corresponding DR3 values. Thus, we select stars that clear the same DR4 cuts described above but which are not in GALAH DR3, with the remaining cuts being:
\begin{itemize}
    \item $\texttt{flag\_sp}=0$
    \item $\texttt{flag\_sp\_fit}=0$ 
    \item $\texttt{flag\_red}=0$
    \item $\texttt{snr\_px\_ccd3}>50$
    \item $\texttt{teff}<6600$ K
\end{itemize}
We also require all of the input DR4 stellar parameters ($T_\textrm{eff}$, $\log g$, $\textrm{[Fe/H]}$, [Ni/Fe]) to have finite values. We thus maintain the requirement for high signal-to-noise spectra and good stellar parameters (both of which are input into the neural network; see Section~\ref{subsec:methods}) and also produce a dataset such that the neural network receives features that are similar to those upon which it was trained. Following these cuts, we have a sample of 118 946 stars in GALAH DR4 which are not in the DR3 cross-match; these are the stars for which we provide new [Eu/H] values.

Following the data selection, these spectra undergo the same process of interpolation onto a uniform wavelength grid, replacement of outlier pixels and scaling to unit variance that was calibrated on the training/validation dataset. Likewise, we select the same wavelength range of $6640$~\AA~to $6650$~\AA~to capture the Eu line, and use the (also rescaled) DR4 stellar parameters $T_\textrm{eff}$, $\log g$, $\textrm{[Fe/H]}$ and [Ni/Fe] as additional input.

\section{Methods}
\label{subsec:methods}

\subsection{Network architecture}

\begin{figure*}
\centering

\begin{tikzpicture}[
    font=\small,
    layer/.style={rectangle, draw, rounded corners, minimum width=1.6cm, minimum height=0.75cm, align=center, fill=bg_io, fill opacity = 0.8},
    conv/.style={rectangle, draw, rounded corners, minimum width=1.6cm, minimum height=0.9cm, align=center, fill=bg_cnn, fill opacity = 0.8},
    fc/.style={rectangle, draw, rounded corners, minimum width=1.6cm, minimum height=0.9cm, align=center, fill=bg_fc, fill opacity = 0.8},
    concat/.style={rectangle, draw, rounded corners, minimum width=0.9cm, minimum height=0.9cm, align=center, fill=bg_concat, fill opacity = 0.7},
    dropout/.style={rectangle, draw, dashed, rounded corners, minimum width=1.5cm, minimum height=0.65cm, align=center, fill= bg_drop, fill opacity = 0.8},
    arrow/.style={-{Latex[length=2mm]}, thick}
]

\node[layer] (spec_in) {Input\\Spectrum\\(159 pixels)};

\node[conv, right=0.5cm of spec_in] (conv1)
    {Conv1D\\kernel = 12\\filters = 16};

\node[conv, right=0.4cm of conv1] (conv2)
    {Conv1D\\kernel = 6\\filters = 32};

\node[dropout, above=0.4cm of conv2] (drop2)
    {Dropout\\(0.14)};

\node[conv, right=0.4cm of conv2] (conv3)
    {Conv1D\\kernel = 3\\filters = 16};

\node[dropout, above=0.4cm of conv3] (drop3)
    {Dropout\\(0.14)};

\coordinate[right=0.5cm of conv3] (flatten);
\node[font=\footnotesize, above=0.14cm of flatten] (flatten2) {Flatten};

\node[fc, right=1cm of conv3] (lin30)
    {Fully\\ Connected\\ 2256 → 30};

\node[layer, below=2cm of conv1] (param_in)
    {Stellar Parameters:\\$T_{\rm eff}$, [Fe/H], $\log g$, [Ni/Fe]};

\node[fc, right=0.5cm of param_in] (fc1)
    {Fully\\ Connected\\4 → 10};

\node[fc, right=0.4cm of fc1] (fc2)
    {Fully\\ Connected\\10 → 20};


\coordinate[right=0.3cm of lin30] (merge0);


\node[concat, font=\small, below=1.2cm of merge0] (concat) {Concatenate\\ (30 + 20 = 50)};


\node[fc, right=0.5cm of concat] (final1)
    {Fully\\ Connected\\50 → 10};

\node[fc, right=0.4cm of final1] (final2)
    {Fully\\ Connected\\10 → 2};

\node[layer, right=0.5cm of final2] (output)
    {[Eu/H]$_{\rm pred}$\\$\sigma_{\rm [Eu/H],pred}$};

\draw[arrow] (spec_in) -- (conv1);
\draw[arrow] (conv1) -- (conv2);
\draw[arrow] (conv2) -- (conv3);
\draw[arrow] (conv3) -- (lin30);

\draw[arrow] (conv2) -- (drop2);
\draw[arrow] (conv3) -- (drop3);

\draw[arrow] (param_in) -- (fc1);
\draw[arrow] (fc1) -- (fc2);

\draw[arrow] (lin30) -| (concat);
\draw[arrow] (fc2) -| (concat);

\draw[arrow] (concat) -- (final1);
\draw[arrow] (final1) -- (final2);
\draw[arrow] (final2) -- (output);

\end{tikzpicture}

\caption{The architecture of our two-branch neural network. Layers are represented by solid boxes, with dashed boxes denoting where dropout is applied and arrows marking the movement of input/output tensors between layers. 1D convolutional layers are labeled as ``Conv1D,'' and linear/fully connected layers are denoted by the ``Fully Connected'' label.}
\label{fig:network_block}
\end{figure*}
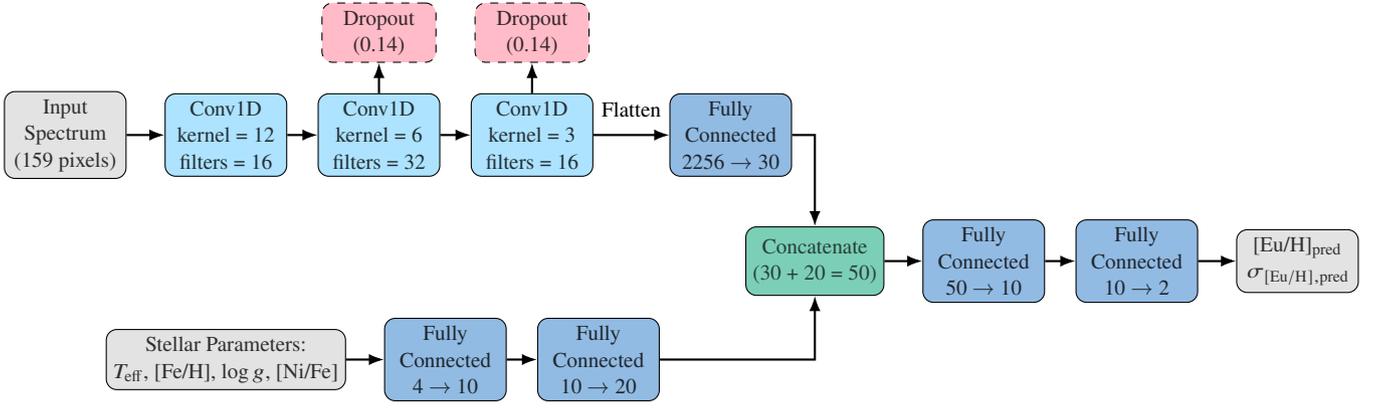

\begin{figure}
	\includegraphics[width=0.9\columnwidth, alt={In the first panel, the CNN-predicted [Eu/H] values closely follow the 1-to-1 line with the GALAH DR3 [Eu/H] values of the validation dataset. The typical uncertainty for a single [Eu/H] value appears to be slightly below 0.1 from the CNN. The root mean squared error (RMSE) value is 0.074, and the R-squared value is 0.88. In the second panel, the absolute value of the CNN prediction residuals are mostly less than 0.2 for [Eu/H] values within 0.5 dex of 0 in either direction, where the bulk of the data lies. There is some clear bias at extreme values, with [Eu/H] abundances below -1 over-predicted by about 0.2 and [Eu/H] abundances near 1 under-predicted by about 0.3. These values are much less common in the validation data.}]{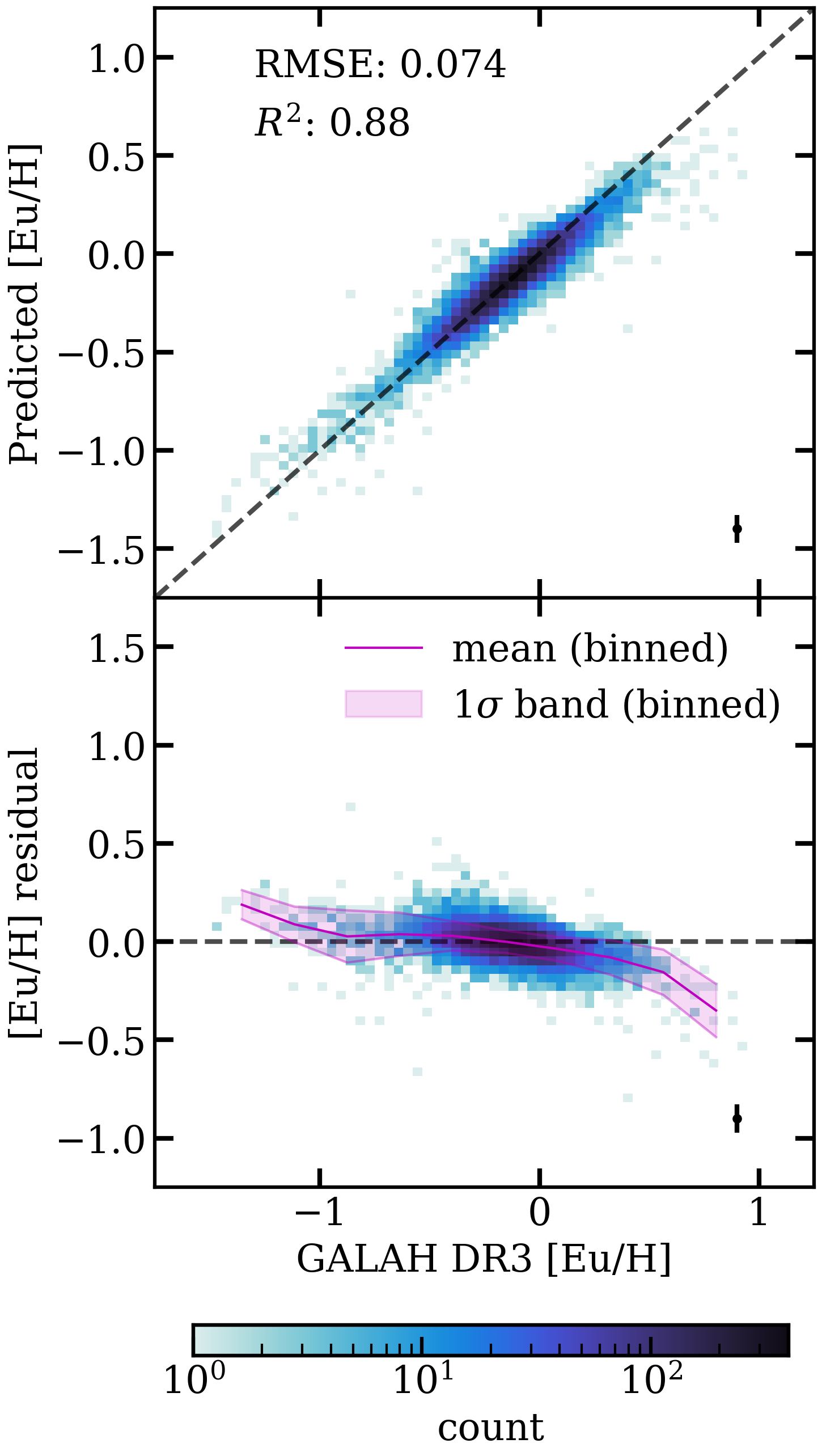}
    \caption{Performance of our model on the validation dataset. A typical (median $\sigma_{\text{y, pred}}$) errorbar for a single prediction is shown in the lower right corner in both subplots for reference. \textit{Top:} A 2D histogram of predicted [Eu/H] values vs. labels (GALAH DR3 [Eu/H] values). Performance metrics RMSE and $R^2$ are listed in the upper left corner. \textit{Bottom:} A 2D histogram of predicted - label residuals. The solid magenta line represents the mean residual value within 10 bins covering the full extent of the label values; the shaded magenta band around it represents 1$\sigma$ intervals. The predicted values show good agreement with labels, although some regression to the mean is still visible in the extreme values.}
    \label{fig:validation_1}
\end{figure}

Our network is implemented in {\tt PyTorch} \citep{Paszke_pytorch}. Similarly to \citet{RAVE_CNN, Guiglion2024, Wang2020, Ambrosch2023}, we build a multi-branch network, working on different types of input separately in several initial layers and concatenating them for final processing stages. As a brief summary, we process interpolated spectra through a branch composed of several 1D convolutional layers to optimally extract information from continuous 1D input, while stellar parameters receive a simpler treatment in a separate branch. We also present a block diagram of our network in Fig.~\ref{fig:network_block}.

The exact architecture of the network is as follows. The branch processing spectra has three consecutive 1D convolutional layers that extract features using decreasing kernel sizes (12, 6, 3), with (16, 32, 16) filters respectively; each with a Rectified Linear Unit (ReLU) activation, which returns the input value for positive inputs, and zero otherwise. To prevent overfitting, we apply dropouts ($=0.14$) after the second and third layers by randomly disabling a fraction of the neurons during training; optimal dropout was determined using the hyperparameter optimization framework {\tt Optuna} \citep{optuna_2019}. The output is then reshaped with a single linear layer into a 30-unit tensor.

The branch processing stellar parameters ($T_{\rm eff}$, [Fe/H], $\log g$, [Ni/Fe]) is a simple two-layer fully connected network with input/output sizes (4, 10), (10, 20), with the output layer being the input into the final, combined block.

The final, combined block takes in 30 units from the spectra and 20 units from the stellar parameters, which are the outputs from their previous respective separate branches. The branch inputs are combined in a fully connected network composed of two layers with input/output sizes (50, 10), (10, 2). For each spectrum, the output consists of a label prediction [Eu/H]$_\text{pred}$ and its reported uncertainty $\sigma_\text{\text [Eu/H], pred}$.

\subsection{Loss}
\label{subsec:loss}
In addition to predicting a $\textrm{[Eu/H]}$ abundance for each spectrum, the network also infers a variance on each prediction, $\sigma_\textrm{[Eu/H], pred}^2$. This is accomplished via the inclusion of a variance term in the loss, which is proportional to the negative log-likelihood of a Gaussian:
\begin{equation}
    \text{loss} = w \left[\frac{(y_{\rm pred} - y_{\rm true})^2 }{\sigma_{y\textrm{, pred}}^2} + \log \sigma_{y\textrm{, pred}}^2\right]
\end{equation}
In the loss, $y_{\text{true}}$ denotes the true value of $\textrm{[Eu/H]}$ from DR3, while $y_{\rm pred}$ is the network-predicted value. This loss form has successfully been used to infer stellar parameters and abundances and their corresponding variances from the \textit{Gaia} low resolution spectra in \citet{Fallows_Sanders_2024, ArdernArentsen_2025, Kane_2025}. $w$ is an optional weighting function that can be applied to alleviate disproportionate representation of data in the training/validation datasets.

In early runs, the prediction of high [Eu/H]$_{\text{true}}$ values -- which had very low representation in the training/validation dataset -- was particularly impacted by regression towards the mean. Considering the high interest in scientific exploration of europium-rich stars, we introduce weighing in the overall loss by the term:

\begin{equation}
w(y_{\text{true}}) = 1 + \alpha \cdot {\rm ReLU}(S(y_{\text{true}}) - 0.3),   
\end{equation}

\noindent where $S$ is the standard scaler to unit variance as implemented in {\tt scikit-learn} \citep{scikit-learn}. For reference, with the scaler calibrated on our training/validation dataset, this weighing means stars with $y_{\text{true}} \leq -0.06$ were not upweighted at all, stars with $y_{\text{true}} = 0.3$ were upweighted by a factor of 2.6, and stars with $y_{\text{true}} = 0.6$ were upweighted by a factor of 4.

We fix $\alpha = 1$, experimentally finding that higher upweighting of this small subset of stars led to strong overfitting, while $\alpha \sim 0.1$ did not have noticeable impact on the predictions. While increased focus on the high-[Eu/H] subset in training was bound to impact the bulk predictions, changes in performance metrics were minimal, with an increase in root mean square error (RMSE) of $0.0007$ and a decrease in $R^2$ score of $0.0025$.

Similar treatment was attempted for metal-poor ([Fe/H] $<$ -1) stars, but was ultimately not retained in the final model. We discuss those experiments and the subsequent treatment of the metal-poor subset in detail in Section~\ref{subsec:metal_poor}.

\subsection{Training}

We train the network for 30 epochs with a batch size of 512, using the {\tt Adam} optimizer \citep{adam} with a weight decay of 0.0093. We start with a learning rate of 0.0008 that is adjusted on plateau by a factor of 0.5, using a patience of 4 epochs. Other settings than listed above are default as defined in {\tt PyTorch} \citep{Paszke_pytorch}. The epoch settings were adjusted manually based on network performance. The optimal learning rate, batch size and weight decay were determined using the hyperparameter optimization framework {\tt Optuna} \citep{optuna_2019}.

For each trained model, the resulting prediction is averaged between 100 samples \citep[following][]{Fallows_Sanders_2024, Kane_2025}. To minimize drops in performance due to random fluctuations at training, we also use ensembling: we train $20$ separate models initialized with different random seeds and average their predictions, resulting in an increase in performance metrics (e.g. RMSE, $R^2$). 

We report all predicted values of [Eu/H], $\sigma_\textrm{[Eu/H]}$ in our catalogue (Section~\ref{sec:catalogue_contents}).

\section{Results}
\label{sec:validation}

In this section, we outline a set of validation tests we performed on our Eu abundances, using both the set of validation data withheld from training and a set of abundances synthesized with the 1D radiative transfer code, \texttt{Korg} \citep{KORG_2023,Korg_2024}.

\subsection{Performance on Validation Dataset}
\label{subsec:validation_performance}

In Fig.~\ref{fig:validation_1}, we demonstrate the overall performance of our model in predicting the labels $y_{\rm true}$ for the validation dataset. We report an RMSE of 0.074 and the $R^2$ score of 0.88. The quality of our predictions of GALAH [Eu/H]$_{\rm DR3}$ labels, made using only DR4 data, sees a dramatic improvement compared to the DR4 dataset. For comparison, [Eu/H]$_{\rm DR4}$ = [Eu/Fe]$_{\rm DR4}$ + [Fe/H]$_{\rm DR4}$ achieves a $R^2$ score of -1.3 as predictor of [Eu/H]$_{\rm DR3}$ labels for the same dataset (after rejecting the 8\% of stars that have {\tt NaN} [Eu/Fe]$_{\rm DR4}$ values).

\begin{figure}
	\includegraphics[width=\columnwidth, alt={The figure shows a 2D histogram of the Kiel (surface gravity versus effective temperature) diagram of stars in our validation dataset, color-coded by the average prediction residual of stars in each bin. Both a main sequence below log(g) of approximately 3.5 and a giant branch at log(g) greater than approximately 3.5 are visible. Exact details on our cuts for dwarfs and giants are given in Section 4.3. Average residuals for giants are uniformly just below 0.1 except for the very edges of the Kiel diagram, where there is slightly more scatter between average residuals of 0.01 and about 0.2. The average residuals on the main sequence is also about 0.1, although there is again slightly more variation here between residuals of 0.01 and about 0.2.}]{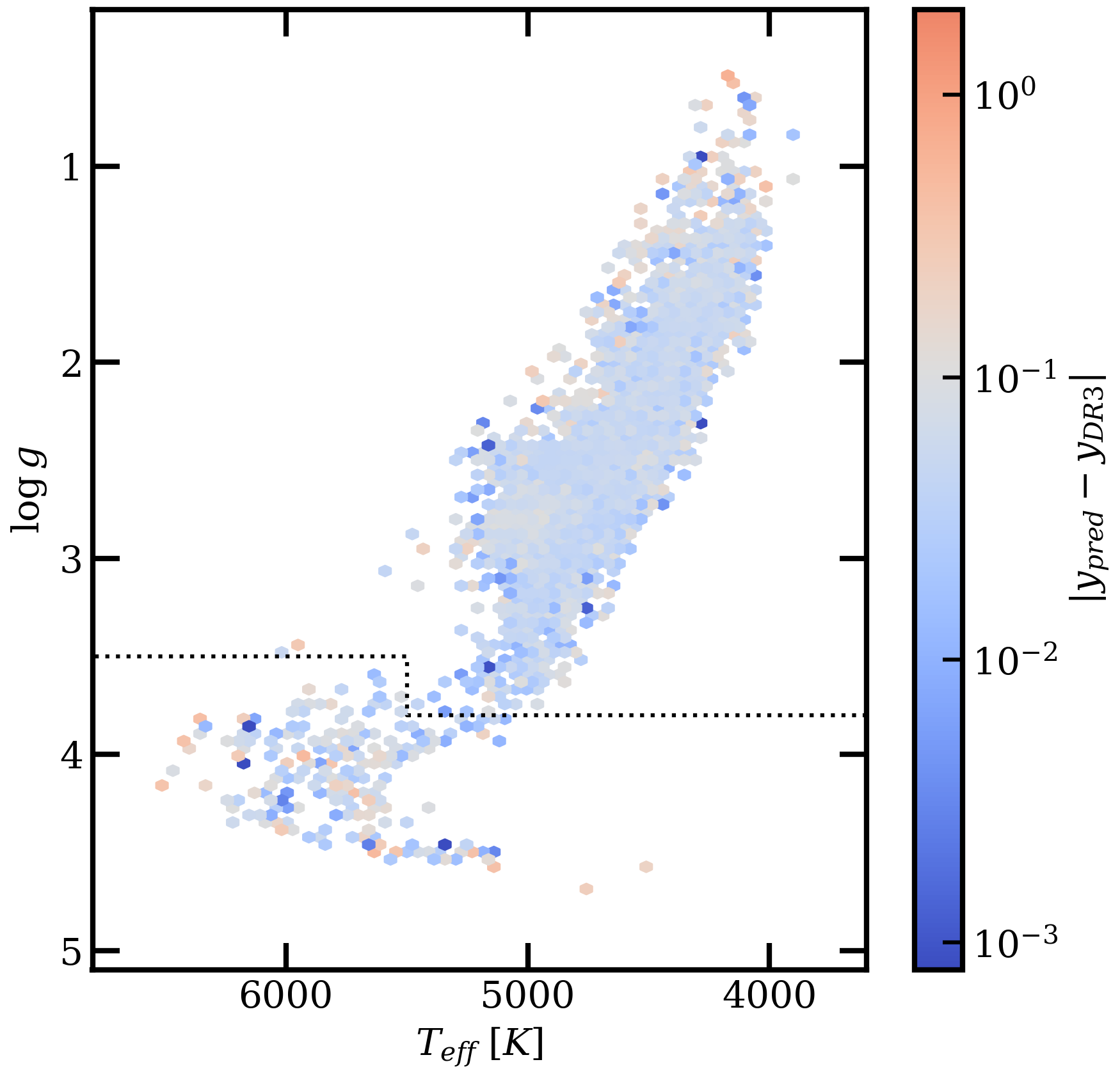}
    \caption{Absolute difference between the prediction ($y_{\rm pred}$) and the label ($y_{\rm DR3}$) within the validation dataset, displayed in 2D hexagonal bins on the Kiel diagram; the dotted black line represents the dwarf-giant boundary (as in Fig.~\ref{fig:train_test}). The model shows consistent good ($<$0.1 dex) performance across a range of stellar parameters except for single outliers (e.g. cold dwarfs, giants on the very low $\log g$ end) whose stellar types are underrepresented in the dataset.}
    \label{fig:validation_kiel}
\end{figure}

In Fig.~\ref{fig:validation_kiel}, we present the performance of our model across different stellar types on the Kiel diagram. The model returns consistently accurate predictions across a range of stellar types, except for several outliers from the giant and dwarf sequences.

\begin{figure}
	\includegraphics[width=1\columnwidth, alt={The left panel show a histogram of z-scores from the validation data, where the z-score is the residual in the CNN prediction divided by the CNN uncertainty. Overplotted are a best-fit Gaussian distribution to the histogram and a comparison Gaussian with a mean of 0 and standard deviation of 1. The histogram of z-scores from the validation data and its assocaited Gaussian fit has a lower mean and lower spread as compared to the model Gaussian with mean 0 and standard deviation 1. The right panel is a line plot showing the confidence interval for the validation data alongisde a 1-to-1 line, which is the nominal confidence interval. The prediction (validation data) confidence interval is uniformly about 5 percent higher than the nominal confidence interval, indicating the the uncertainties inferred by the CNN are slightly over-predicted, although generally the number of stars whose predictions fall within their projected uncertainties is in relatively good agreement with a Gaussian.}]{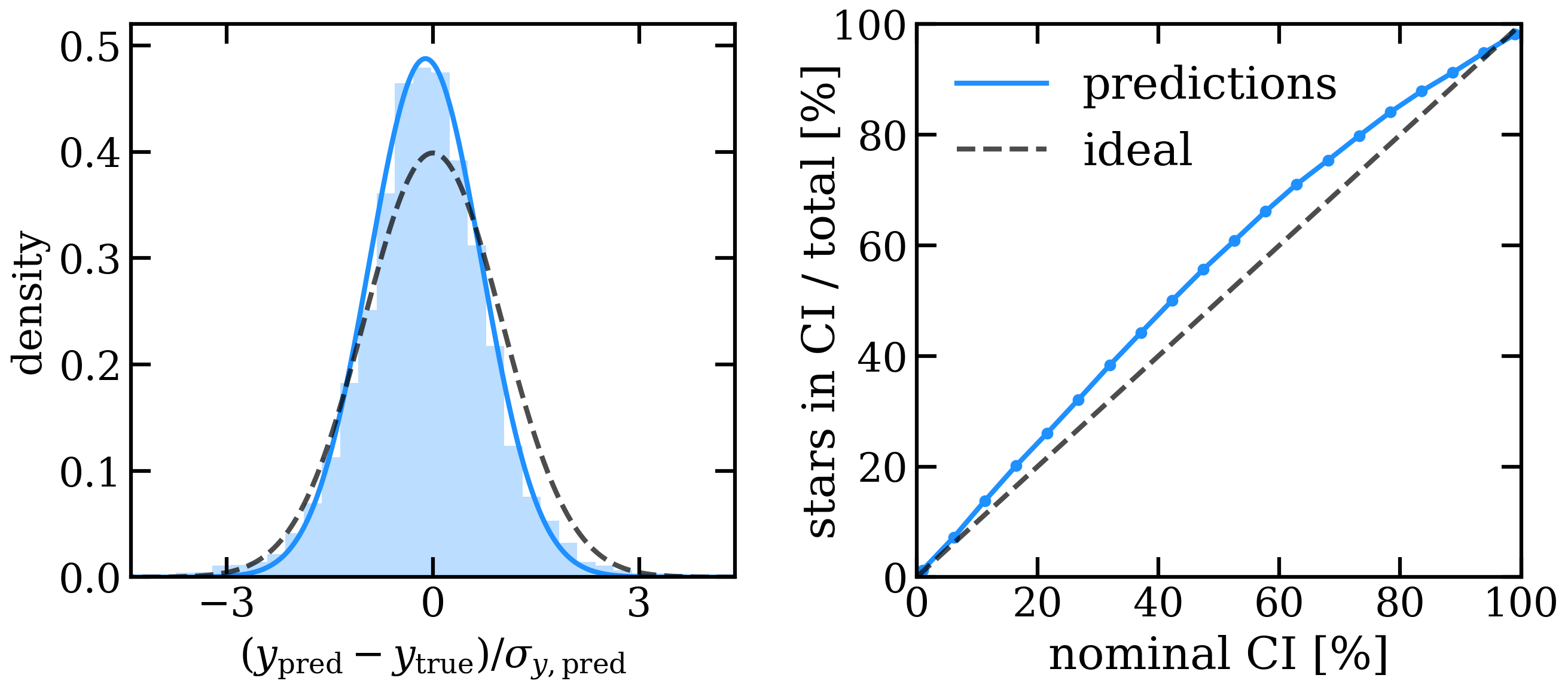}
    \caption{Analysis of the agreement of uncertainties $\sigma_{y, \ {\rm pred}}$ reported by our network for the validation dataset and a theoretical model of Gaussian noise. \textit{Left:} $z$-score histogram of network predictions (light blue). The ideal distribution (dark grey dashed line) is a Gaussian ($\mu=0$, $\sigma=1$). The solid blue line represents a Gaussian ($\mu=-0.11$, $\sigma=0.82$) fit to the histogram; as there are $\approx 10$ extreme outliers in the $x$-axis that strongly impact the distribution fit, a robust ($\mu$, $\sigma$) fit was obtained using only (25, 50, 75)$^{\rm th}$ percentile values. \textit{Right:} A comparison of nominal and empirical confidence intervals. The $x$--axis represents a theoretical confidence interval used to calculate $n$, while the $y$--axis represents the actual fraction of stars whose $y_{\rm true}$ is covered by $y_{\rm pred} \pm n \sigma_{y, \text{pred}}$. The ideal case is represented by a 1:1 dark grey dashed line. The empirical curve calculated for 20 different confidence intervals spanning from 1\% to 99\% (blue points) using network predictions is plotted in solid blue.}
    \label{fig:variance_distr}
\end{figure}

In addition to predicting the [Eu/H] label, our model also assesses the prediction uncertainty $\sigma_{\rm [Eu/H]}$. In Fig.~\ref{fig:variance_distr}, we use the $z$-score distribution to verify that the predicted--true residuals are well-modelled by Gaussian noise scaled by $\sigma_{\rm [Eu/H]}$. We also show that the number of stars whose predictions fall within network-reported confidence intervals show good agreement with the theoretical model. Our model shows a very slight negative bias in label predictions ($\bar{z} = -0.11$) and a slight overestimation of uncertainties ($\sigma_z = 0.82$).

\begin{figure}
	\includegraphics[width=\columnwidth, alt={The figures shows a series of histograms showing the network's inferred standard deviations on the [Eu/H] predictions for the validation data versus GALAH DR4 [Fe/H], [Ni/Fe], effective temperature, surface gravity, and the GALAH DR3 [Eu/Fe]. The inferred standard deviations are almost all between approximately 0.05 and 0.1 and are mostly uniform relative to these other variables. Notably, there is a dense group of validation stars at a surface gravity of 2.5, which is likely red clump stars. Standard deviations also increase to slightly above 0.1 for stars with [Fe/H] less than -1 and stars hotter than about 6000 K, although there are very few of these stars in the data.}]{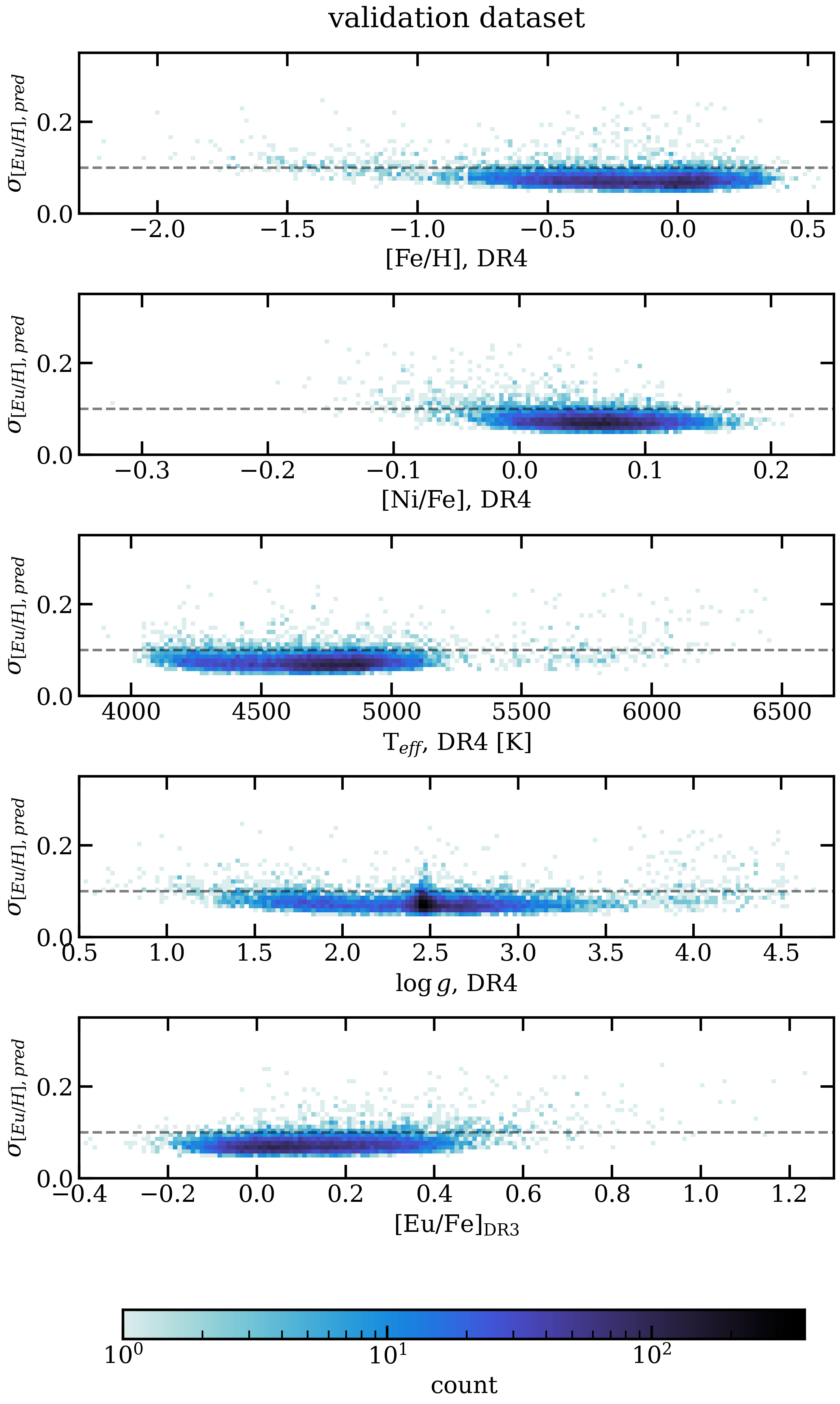}
    \caption{$\sigma_{[\rm Eu/H]}$ predicted by our CNN for the validation dataset as a function of stellar parameters (from top to bottom: stellar parameter branch input: [Fe/H], [Ni/Fe], $T_{\rm eff}$ and $\log g$, as reported in GALAH DR4, and [Eu/Fe] as reported in DR3 and used for constructing the label). Color coding represents star counts per bin. The grey dashed line marks the 0.1 dex level for reference. Our model shows near-constant, $\lesssim$0.1 dex uncertainties across a range of stellar parameters that is well-covered in the training/validation dataset. The most notable drop in confidence of model predictions happens for metal-poor ([Fe/H]$<-1$) stars, which are not only underrepresented in the training/validation dataset but also have weaker spectral lines. Other stellar parameter ranges that are not well-covered, e.g. the low $\log g$ edge or the high [Eu/Fe] edge, also have decreased confidence of model predictions. (Note: 4 stars (0.04\% of the validation dataset) are outliers with $\sigma_{\text{[Eu/H], pred}}>0.35$ and were not included here for clarity of the figure.)}
    \label{fig:val_variance_params}
\end{figure}

In Fig.~\ref{fig:val_variance_params}, we analyze these uncertainties as a function of stellar parameters: the network input parameters ($T_{\rm eff}$, [Fe/H], $\log g$, [Ni/Fe]) and the DR3 [Eu/Fe] values. $\sigma_{\rm [Eu/H]}$ increases for poorly-covered stellar parameter ranges; this behaviour is most notable for metal-poor stars, which are both rare and have weak spectral lines. However, we find consistently high-precision ($\lesssim$0.1 dex) predictions across a range of stellar parameters covered in the training/validation datasets, and conclude $\sigma_{\rm [Eu/H]}$ is not strongly impacted by systematic effects in the validation dataset.

\subsection{Feature Importance}
\label{sec:feature_importance}
We perform a standard permutation feature importance analysis. We first pre-train the network ensemble using the full dataset described in Section \ref{subsec:test_train}. In addition to the standard atmospheric parameters ($T_{\rm eff}$, [Fe/H], $\log g$), we also include further relevant parameters to explore whether the input should be expanded. Those parameters are: microturbulent velocity $v_{\rm mic}$, nickel abundance [Ni/Fe], and rotational velocity $v \sin i$; we use the stellar labels published in GALAH DR4 for all of them. We change the parameter branch input size accordingly to 6. Due to this increase, network training takes longer to converge; we extend training to 50 epochs. All other hyperparameters are as described in Section~\ref{subsec:methods}.

We then partially corrupt the validation dataset by randomly shuffling each feature\footnote{While we implement the feature importance analysis as a custom function due to the varying feature sizes, the workflow is similar to that outlined here: \url{https://scikit-learn.org/stable/modules/permutation_importance.html}.}. For each corrupted dataset, we make predictions and compute performance metrics as in Section \ref{subsec:validation_performance}.

\begin{figure}
    \includegraphics[width=\columnwidth, alt{The figure shows a normalized spectrum in our 10 angstrom window with regions color-coded by their feature importance. The most important region in teh spectrum is the Eu line, followed buy the nickel line and the iron line at 6646.93 angstroms. The iron line at 6648.08 angstroms is less important, and the edges of the wavelength region have less importance than the typical performance scatter between an ensemble of networks. There is an inset plot showing labels for input stellar, also color-coded by feature importance. In order of highest to lowest feature importance, those labels are [Fe/H], log(g), T, [Ni/Fe], micro turbulence, and v sin(i). [Fe/H] has slightly higher importance than the Eu line in the spectrum, and micro turbulence and v sin(i) have lower importance than the typical performance scatter between networks.}]{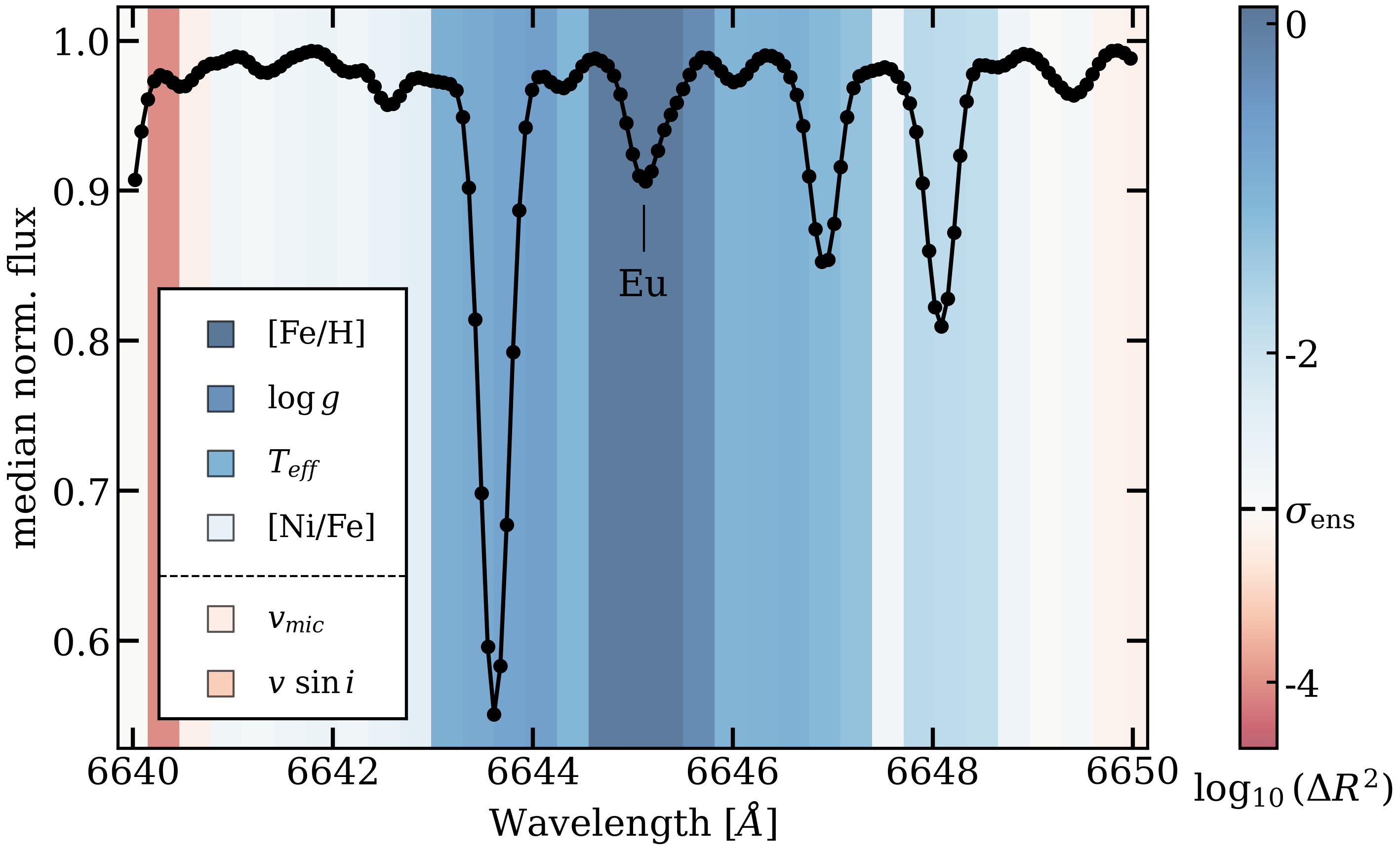}
    \caption{Results of permutation feature importance analysis on the validation dataset. The stellar parameters (inset) and the normalised spectrum (main plot) share the same colour scale; colours represent the decrease of $R^2$ score between the original prediction and the prediction made using partially corrupted data. Dark blue shades signify highest importance, while warm shades signify negligible importance; the colour scale zero-point (light gray) is calibrated to the level of scatter between individual networks in the ensemble for comparison, also indicated in the colorbar by a dashed black line. For the stellar parameters ($T_{\rm eff}$, $\log g$, [Fe/H], $v_{\rm mic}$, [Ni/Fe], $v \sin i$), each parameter is treated as a single feature, and their importance is overplotted in the inset (lower left) in decreasing order. For the normalised flux, a `feature' is defined by a 20-gridpoint sliding window (moving by 5 gridpoints), chosen to approximately match typical linewidths. Feature importance is represented by the colour of the background around the midpoint of the window. The median normalised spectrum of the training/validation dataset (median stellar parameters: $T_{\rm eff}$ = 4736 K, $\log g$ = 2.47, [Fe/H] = -0.18, [Ni/Fe] = 0.06) is overplotted for reference. The [Fe/H] parameter has the highest importance for network predictions, closely followed by the Eu line. The neighbouring Ni and Fe lines and the stellar parameters $\log g$, $T_{\rm eff}$ are also of high importance. [Ni/Fe] and most continuum/weak line regions have low, but still statistically significant importance. Edges of the wavelength range and stellar parameters $v_{\rm mic}$, $v \sin i$ have negligible impact on network predictions. Stellar parameters under the dashed line in the inset were not included in the final network described in Section~\ref{subsec:methods}.}
    \label{fig:feature_importance}
\end{figure}

Flux values from continuous gridpoints are highly correlated and, therefore, impractical as single features (see Fig.~\ref{fig:feature_importance}). We instead apply a sliding window of 20 gridpoints (slightly larger than the typical line width), moving by 5 gridpoints between runs. (To cover the entire wavelength range, we allow shorter windows at the edges.) We treat stellar parameters as single features.

For all features, we assess their importance based on the mean drop in performance metrics in predictions where they are corrupted (illustrated in Fig.~\ref{fig:feature_importance} for $R^2$ scores). We decided to include features in the final model if it led to improvement in performance metrics at least as high as the level of scatter between individual networks in the ensemble, which we use as a reference value. As expected in the linear regime of the curve of growth, the Eu line and [Fe/H] are the most important features ($\Delta R^2 \sim 1.0$), followed by the Ni and Fe lines directly bordering Eu and stellar parameters $T_{\rm eff}$ and $\log g$ ($\Delta R^2 \sim 0.1$). The regions of the spectrum containing other weak lines and continuum, as well as [Ni/Fe], have lower but statistically significant importance ($\Delta R^2 \sim 0.01$). The importance of added parameters $v_{\rm mic}$, $v \sin i$ is lower than the noise level between networks initialised with different random seeds ($\sigma_{\rm ens} = 0.0011$). Keeping input size (and thus, the stellar parameter branch size) limited serves to optimize computational resources needed for training and predictions and limit overfitting. Initially, we trained our model on ($T_{\rm eff}$, [Fe/H], $\log g$) only; after extending the parameter set with [Ni/Fe], model performance became comparable to the 6-parameter network, while requiring less resources.

We note that while the network trained on the full dataset heavily relies on [Fe/H] information, we also tried training the final network with selected features missing entirely. In this experiment, the network adapts to instead learn the missing information (e.g. metallicities) from other input. While removing [Fe/H] from the training dataset increases scatter in predictions, the drop in performance is significantly less dramatic in this case than with corrupted data without new training (e.g. $\Delta R^2 = 0.051$ for missing vs. $\Delta R^2 = 1.608$ for corrupted [Fe/H]). This validates the use of our approach also in cases where stellar parameters are not available or reliable.

In summary, in this section we analysed which input is most impactful for [Eu/H] predictions. We decided to limit our input to the parameters that significantly improve prediction quality. We also tested that when deprived of important information like metallicities, the network learns to use other input in its place. In addition to guiding the construction of the network, this exercise serves as a soundness check for our model and a possible guide for future studies.

\subsection{Data Quality Cuts}
\label{subsec:data_cuts}

\begin{figure}
    \includegraphics[width=0.8\columnwidth, alt={In the first panel, the spectra that pass the continuum flux cut clearly sit at about a normalized flux of 1, except for the distinct nickel, iron, and europium lines. By contrast, the spectra excluded by this cut fall at much lower normalized fluxes, with some having continua only about 0.1 below the expected normalized flux of 1 and others being as low as 0.5 or lower. In they spectra excluded by the continuum cut, the Eu feature appears distorted, and there are molecular features visible in addition to the Fe, Ni, and Eu lines. The second panel shows the distribution of continuum fluxed=s, with any spectra with continuum fluses below 0.93 or 1.07 excluded. Values between 0.93 or 1.07 are the most common, with over 100,000 stars falling within this range. Several thousand stars have continuum fluxes below 0.93, with the frequency of stars falling with decreasing contiuum fluxes. Only a few stars have continuum fluxes greater than 1.07.}]{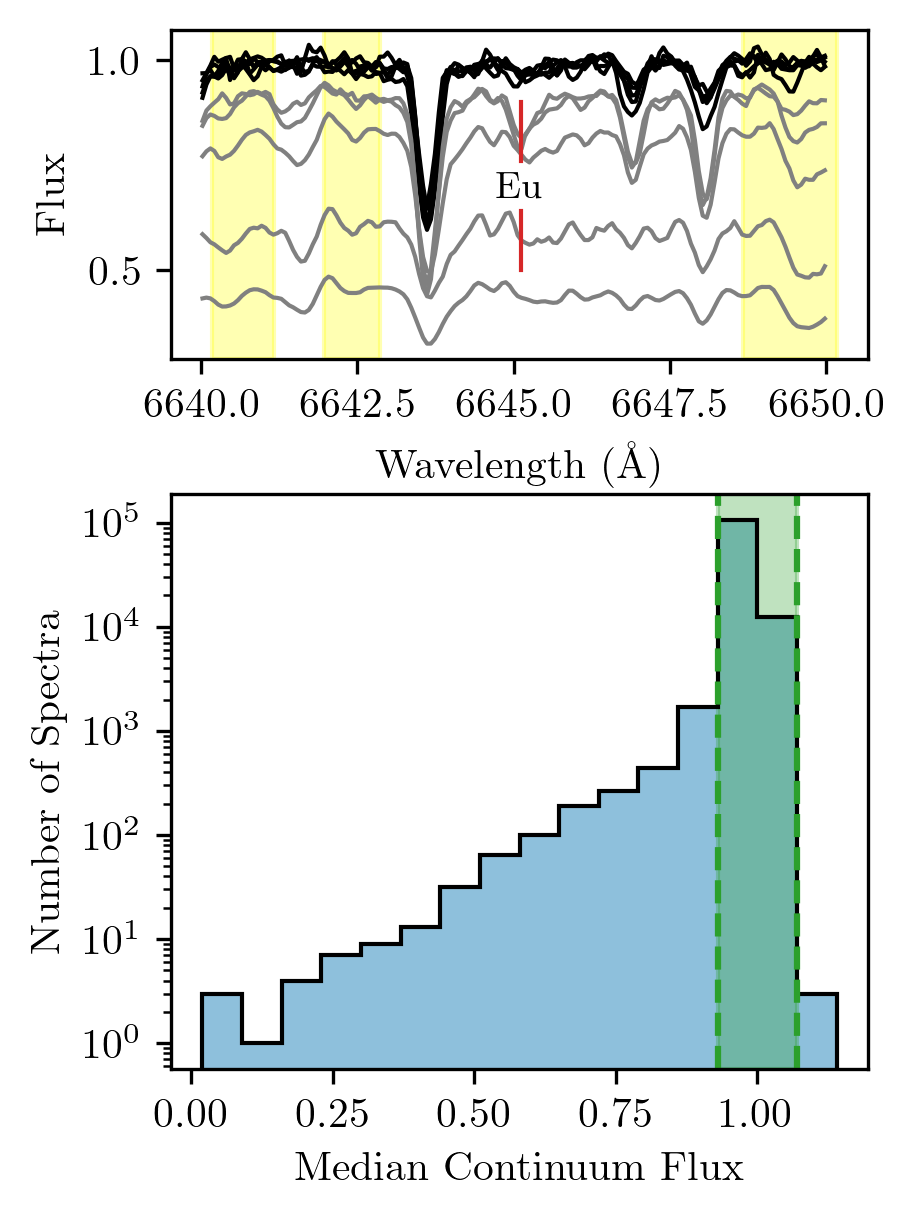}
    \centering
    \caption{Top: A selection of GALAH DR4 spectra in the wavelength region centered on the Eu line, illustrative of the continuum flux selection used in this work. The yellow windows mark the regions used to calculate the median continuum flux. The black spectra are those that clear our requirement of having a median normalized continuum flux in the marked windows $>0.93$ and $<1.07$; the grey spectra do not meet this continuum flux requirement and are flagged. The Eu line's location is labeled and marked in red for reference. Bottom: A histogram of the median fluxes of GALAH DR4-exclusive (see Section~\ref{subsec:data}) stars in the continuum windows marked in the top panel. The green, dashed lines mark our cutoff at a median normalized continuum flux between $0.93$ and $1.07$; outside the shaded band defined by this cutoff spectra are flagged with \texttt{continuum\_flux\_flag}.}
    \label{fig:continuum_flux}
\end{figure}

To ensure the most reliable Eu abundances possible for community use, we provide here a set of recommended data cuts and quality flags to apply to the catalogue. These cuts are motivated by our experimentation with the predictions, validation performance, network-generated uncertainties, and science testing on the DR4-exclusive data. For the convenience of catalogue users, we provide three new data quality flags in our catalogue of abundance predictions. These flags are \texttt{stellar\_params\_flag}, \texttt{continuum\_flux\_flag}, and \texttt{line\_depth\_flag}. Each of these should be set to $0$ to ensure there are no issues flagged and will appear as $1$ if a problem has been identified. Applying the final flag option in the table, $\texttt{eu\_flag}=0$, will ensure that all three of these flags are $=0$ and that you are using our best sample of [Eu/H] abundances.

It is also useful at times to distinguish between dwarfs and giants for the purposes of quality cuts, and so for this work we make a cut consistent with the work of \citet{Borisov2022}. In particular, giants are defined as stars with $T_\mathrm{eff}\geq5500$ and $\log g<3.5$ or $T_\mathrm{eff}<5500$ and $\log g<3.8$. Conversely, dwarfs are those stars with $T_\mathrm{eff}\geq5500$ and $\log g\geq3.5$ or $T_\mathrm{eff}<5500$ and $\log g\geq3.8$. These cuts are also visualized via the black dotted lines in Fig.~\ref{fig:train_test}.

\begin{figure*}
    \includegraphics[width=2\columnwidth, alt={The left panel shows six example spectra with Eu line depth values ranging from -0.02 to 0.07. The spectrum with the lowest line depth has no visible Eu line at all. Then, with increasing Eu line depth the Eu line appears and grows stronger. In teh spectrum with the highest line depth of 0.07, the flux depth of the Eu line does appear to be approximately 0.07. The continuum windows are immediately to the left of the nickel line at 6643.63 angstroms, in between the nickel line and the europium line at 6645.11 angstroms, and to the right of the europium line but prior to the iron line at 6646.93 angstroms. The europium flux window covers the entirety of the europium feature at 6645.11 angstroms. The other two panels show histograms of Eu line depths from our training and validation data and from the DR4-exclusive data, in separate panels for dwarfs and giants. Among the giants, there are fewer than 1000 stars fall below our line depths cut of 0.01 in the training and validation data. By contrast, over 1000 stars in teh DR4-exclusive data are excluded by the cut, demonstrating that these weak-line giants were not very well-represented by the training data. Most giants have line depths greater than 0.01 in both the training and validation and DR4-exclusive data, with a few even having line depths greater than 0.1. By contrast, many of the dwarfs have weak line depths, with essentially none having line depths greater than 0.05. In the training and validation data, fewer than 100 dwarfs have line depths lower than 0 as opposed to several hundred in the DR4-exclusive data. Tens of thousands of the DR4-exclusive dwarfs are excluded by our line depth cut at 0.02 as opposed to only about 1000 dwarfs from the training and validation data falling beneath this cut, again demonstrating that the weak-line dwarfs often do not have Eu abundances in GALAH DR3 and thus are not represented in teh training data.}]{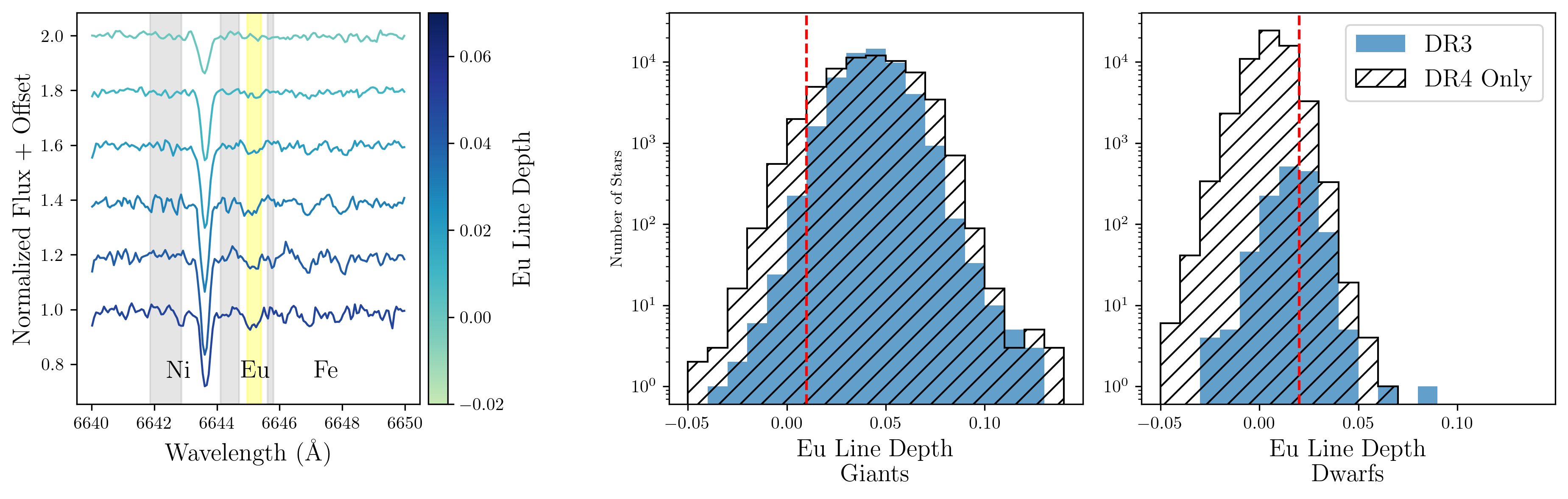}
    \caption{Left: A series of dwarf star spectra with varying Eu line depths indicated by the colorbar. A small vertical offset of 0.2 is applied to the normalized flux for easier visualization. Highlighted in yellow is the region of the spectrum used to calculate the flux in the Eu line, and highlighted in gray are the regions used to calculate the continuum flux. Middle: The distributions of Eu line depth for giant stars in our GALAH DR3 training and validation samples (blue) and the GALAH DR4-exclusive dataset (black hatched). Right: The same as the middle, but now showing the distribution of Eu line depth for dwarfs in each of the two samples. In the middle and right panels, the red, dashed vertical lines mark our imposed cuts on the required Eu line depth at $>0.01$ for giants and $>0.02$ for dwarfs.}
    \label{fig:eu_line_strength}
\end{figure*}

\texttt{stellar\_params\_flag}: As described in Section~\ref{subsec:data}, we apply a cut on the effective temperature $T_\textrm{eff}<6600$~K, which affects exclusively the dwarf sample. This cut is motivated by the fact that these stars are a.) uncommon in our training data, and thus difficult to robustly make predictions for, and b.) more importantly, appear to have spuriously high [Eu/H] abundances in GALAH DR3, leading to seemingly non-physical high [Eu/H] abundance predictions in the DR4-exclusive data set. Unlike the recommended quality cuts outlined next, this cut is applied to the training and validation set itself, such that the CNN is not trained on stars with very high Eu abundances. In addition to the universal $T_\mathrm{eff}<6600$~K cut applied to all data, we also strongly recommend removing cool stars by requiring $T_\mathrm{eff}>4000$~K. This cut is primarily motivated by the fact that such stars are essentially nonexistent in the training and validation data, meaning that the network has to extrapolate to make predictions for such stars. Examinations of the CNN's predictions for these cool stars showed them to have unusually low [Eu/H] abundances, thus motivating their removal. Moreover, many stars with $T_\mathrm{eff}<4000$~K have continuum fluxes $\ll 1$ in our selected wavelength region, as described further in the section for the \texttt{continuum\_flux\_flag}, which results in anomalous [Eu/H] predictions.

\begin{figure*}
    \includegraphics[width=2\columnwidth, alt={The left two panels show scatter plots of the GALAH DR3 or CNN predicted [Eu/H] versus the Korg-derived [Eu/H], respectively, with a 1-to-1 line overplotted for reference. The mean absolute difference between GALAH DR3 and Korg is 0.0963, and between the CNN and Korg it is 0.0923. [Eu/H] values in both panels range between approximately -1.1 and 0.6. Both GALAH DR3 and the CNN have [Eu/H] values approximately 0.1 dex lower than Korg on average, although the CNN also has about 15 stars with [Eu/H] predictions approximately 0.1 dex higher than Korg. Such a group is not present in the GALAH DR3 comparison. The typical CNN-inferred uncertainty in [Eu/H] is about 0.1 and much smaller than 0.1 from korg. The right panel shows three example GALAH spectra in our wavelength region and their associated fits with Korg. Two of the three spectra have good Korg fits to the Eu line; the third has an additional Fe line at 6645.37 angstroms which is blended with the Eu line and is not well-fit by Korg. The rest of the spectrum regions, especially the continua, are well fit by Korg with the exception of the iron line at 6646.93 angstroms, which is not fit in any of the three spectra.}]{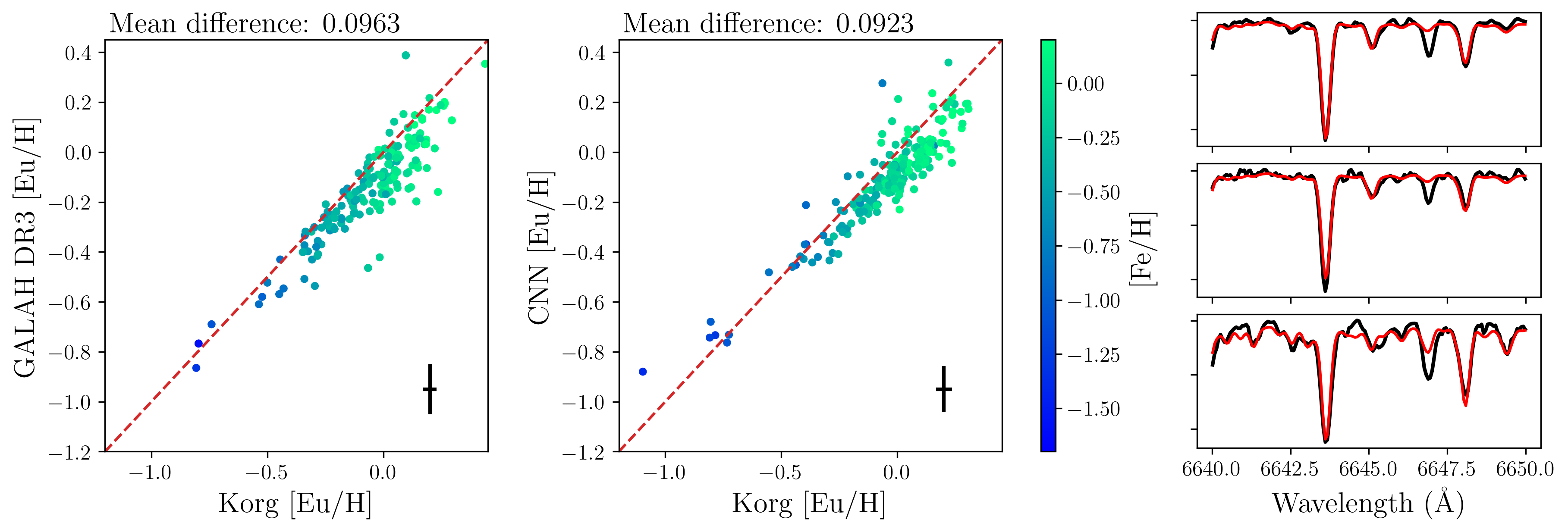}
    \caption{The \texttt{Korg} comparison of [Eu/H] abundances for a set of giant stars which pass our quality cuts. On the left, a comparison of $200$ stars' GALAH DR3 and \texttt{Korg}-derived [Eu/H] values, color-coded by [Fe/H]. The middle panel shows CNN versus \texttt{Korg} [Eu/H] values for a set of $200$ DR4-exclusive giants. In the left and middle panels, the error marker shows the average uncertainties. The mean absolute difference between the two sets of [Eu/H] values is also shown for both panels. The right three panels show three sample spectra from the DR4-exclusive dataset in our selected wavelength region. The black lines mark the observed GALAH spectrum, while the red line is the best fit flux from \texttt{Korg}. Note that the atomic data does not accurately reproduce the observed flux for two of the Fe lines.}
    \label{fig:korg_validation_giants}
\end{figure*}

In addition to the temperature cuts, we also suggest using only stars with metallicities [Fe/H]~$>-1.7$. This recommendation is not guided by network performance but is instead motivated by the fact that the Eu line at $~6645.11$~\AA~grows uniformly very weak at metallicities $<-1.7$ \citep[many studies will use the Eu feature at $4129$~\AA~and others in the UV for very metal poor stars, see e.g.,][among others]{RPA_DR1,RPA_DR4_2020}. Requiring $\texttt{stellar\_params\_flag}=0$ in our catalogue will remove stars not meeting these temperature and metallicity requirements.

\texttt{continuum\_flux\_flag}: Examination of the GALAH DR4 spectra reveals that a small but not insignificant fraction of stars in our dataset ($\approx 2\%$) have continua in our wavelength region of $6640$~\AA~to $6650$~\AA~that are not consistent with $1$. These targets are predominantly cool stars ($T_\mathrm{eff}\lesssim4000$) with strong CN and TiO molecular features that affect the normalization of the continuum in the wavelength range used here. To systematically identify and remove these spectra from our data, we identify three windows that exclude the Eu, Ni, and Fe lines in this wavelength region. These windows are $6642-6643$~\AA, $6643.8-6644.7$~\AA, and $6650.5-6652$~\AA~and are marked in Fig.~\ref{fig:continuum_flux} as the yellow regions. Although some molecular features exist within these windows, because the spectra with continuum fluxes $\ll1$ are from cool stars with strong molecular absorption, including fluxes in the region of such features is acceptable for our purposes here. The top panel of Fig~\ref{fig:continuum_flux} provides illustrative examples of spectra with and without normalization issues. We flag all spectra where the median continuum flux within these windows is not $>0.93$ and $<1.07$ with $\texttt{continuum\_flux\_flag}=1$. In total, 2816 stars from our DR4-exclusive prediction dataset (2.4\% of the dataset) are flagged in this process.
Among the DR4-exclusive stars flagged as having continua outside our target range, the mean $T_\mathrm{eff}$ is $3836$~K, as compared to a mean $T_\mathrm{eff}$ of $5284$~K among the stars that clear the cut. We thus note that our recommended $T_\mathrm{eff}<4000$~K cut within \texttt{stellar\_params\_flag} will remove most (85\%) of these stars. Applying a cut on \texttt{continuum\_flux\_flag} only removes an additional $\approx400$ stars with respect to a cut on \texttt{stellar\_params\_flag} and thus does not significantly limit the total sample size.

\texttt{line\_depth\_flag}: To supplement our catalogue of predicted [Eu/H] abundances, we also provide an approximate measure of the Eu line depth for each star. These values are intended only as approximations of the Eu line depth and can be used to estimate the detectability of europium abundances; for sufficiently weak lines, the abundances can be considered to be upper limits. Most importantly, these line depth estimates are used to make appropriate data cuts such that the DR4-exclusive data -- particularly for the dwarfs -- is well-represented by the training and validation dataset. This is the primary function of these line depth values, which are also provided in the catalogue in the \texttt{line\_depth} column alongside the \texttt{line\_depth\_flag}.

The Eu line depth is estimated as follows. We select several windows covering the continuum (excluding the molecular features, as best as is possible) and one window focused on the Eu line. The Eu line window is asymmetrical to capture the region of the feature with the strongest absorption, taking into account the effects of hyperfine structure and isotopic splitting on the Eu line \citep{Lawler_2001, Sneden_2008}. The continuum and Eu line windows are marked in Fig.~\ref{fig:eu_line_strength}. The Eu line depth is then simply calculated as the mean flux in the (continuum-normalized) continuum windows minus the median flux in the Eu line window.

Fig.~\ref{fig:eu_line_strength} illustrates our Eu line depth estimation alongside the line depth distributions of dwarf and giant stars, shown separately. Most important to note from the left panel, which shows examples of dwarf star spectra with a range of Eu line depths, is that the visibility of the Eu line appears to increase with greater values of Eu line depth. In particular, stars with a $\texttt{line\_depth}\approx0$ have no visible Eu line in their spectra. These line depth measurements are especially important for the dwarf stars, as there are many more dwarfs in the DR4-exclusive dataset than in the DR3-DR4 cross-match dataset used in training and validation. Notably, as shown in the middle panel of Fig.~\ref{fig:eu_line_strength}, the line depth distributions between these two samples are markedly different. The DR4-exclusive dwarfs extend to smaller line depths than their DR3 counterparts. To select only dwarfs which are both well-represented in our training data and which have visible Eu lines, we require $\texttt{line\_depth}>0.02$ for all dwarfs.
The distributions of line depths for giants between GALAH DR3 and DR4-exclusive stars also differ slightly (see middle panel of Fig.~\ref{fig:eu_line_strength}), but not to the extent of the dwarf stars. For this reason, we impose a milder cut of $\texttt{line\_depth}>0.01$ on giants, which our testing has shown to improve performance.

To conclude, our full list of recommended data cuts, which can all be applied in the catalogue by requiring $\texttt{eu\_flag}=0$, are
\begin{itemize}
    \item $4000 < T_\mathrm{eff} < 6600 $ K
    \item $\mathrm{[Fe/H]}>-1.7$
    \item Continuum flux $>0.93$ and $<1.07$
    \item For dwarfs, $\texttt{line\_depth}>0.02$
    \item For giants, $\texttt{line\_depth}>0.01$
\end{itemize} 

\subsection{Performance on a Secondary Set of Synthesized Abundances}
\label{subsec:korg}

As a secondary check, we also measured the Eu abundances for 800 additional (i.e., not part of the training/validation dataset) stars exclusive to GALAH DR4 by synthesizing their spectra using \texttt{Korg} \citep{KORG_2023,Korg_2024}. The stars in the secondary validation sample are split between giants and dwarfs with strong Eu lines per Section~\ref{subsec:data_cuts} and are selected to span a range of stellar parameters representative of the overall dataset. Each sample of 400 giants or dwarfs is further divided into two subsets. 200 stars have GALAH DR3 Eu measurements, which we use to benchmark the performance of our \texttt{Korg}-derived values. The further 200 stars have no DR3 [Eu/H] abundances and thus are unique from the validation sample in Section~\ref{subsec:validation_performance}. We use these additional 200 stars with no DR3 [Eu/H] as a further validation of our CNN abundances.

To perform the fitting, we adopt the GALAH line list \citep{GALAH_DR3} and use the full 10~\AA~window of our spectra surrounding the Eu line for the \texttt{Korg} fit, which we found to provide a better the continuum fit and thus to improve agreement between the GALAH DR3 [Eu/H] values and our own for the purposes of this comparison (although note that it is only the Eu feature itself being fit, as other input stellar parameters and abundances are held fixed at their GALAH DR4 values). Similar to what was done in GALAH, we use 1-dimensional MARCS model atmospheres \citep{Gustafsson_2008} which are plane-parallel for stars with $\log g>3.5$ and spherical for stars with $\log g<3.5$. The abundance of $\alpha$ elements in the model atmosphere are adopted in accordance with GALAH DR3 \citep{GALAH_DR3}, such that stars with [Fe/H]~$\leq-1$ have $\mathrm{[\alpha/Fe]}=0.4$, stars with [Fe/H]~$\geq0.0$ have $\mathrm{[\alpha/Fe]}=0.0$, and stars with intermediate metallicities have $\alpha$-element abundances linearly interpolated between these two values. In addition to the spectrum itself, we also provide \texttt{Korg} with the error spectrum as well as the GALAH DR4 values of $T_\mathrm{eff}$, $\log g$, [Fe/H], $v_\mathrm{mic}$, and $v \sin{i}$. Specifying other elements, such as Ni and Ti, for the \texttt{Korg} input appeared to have no substantial effect on the results. As for GALAH DR3 and DR4, we assume local thermodynamic equilibrium when synthesizing the Eu line region with \texttt{Korg}.

We show the results comparing [Eu/H] values for both GALAH DR3 and \texttt{Korg} and, separately, the CNN's predictions and \texttt{Korg} for giant stars passing our quality cuts in Fig.~\ref{fig:korg_validation_giants}. Both GALAH DR3 and our CNN have values that coincide with \texttt{Korg} to within $\leq0.1$ on average. Although GALAH DR3 and \texttt{Korg} do not agree perfectly on Eu values, there is nonetheless a relatively good consistency between the two, with $\sim64\%$ of our GALAH DR3 comparison stars having [Eu/H] values $\leq0.1$ away from the values we fit with \texttt{Korg\footnote{Systematic differences between GALAH DR3 and our \texttt{Korg} abundances are not surprising given that DR3 abundances were determined with SME \citep{Piskunov2017}, a different radiative transfer code.}}. The standard deviation of the errors between \texttt{Korg} and GALAH DR3 are also low at a value of $0.083$, indicating relatively good precision. This consistency suggests that spectral synthesis with \texttt{Korg} is a sufficient secondary validation for our CNN abundances. Among the CNN comparison for giants, $\sim59\%$ of the 200 stars agree with \texttt{Korg} [Eu/H] values to within $<0.1$ and $\sim96\%$ of the stars agree within $<0.2$; the standard deviation of the errors is $0.057$. Interestingly, there is similar behavior among both the GALAH DR3 and CNN [Eu/H] values, with both being systematically slightly lower than \texttt{Korg}, especially at higher metallicities. This is promising given that GALAH DR3 is our true target for comparison for the CNN.

Upon comparison with the observed GALAH spectra, it is worth noting that the \texttt{Korg} fits are \emph{not} universally perfect. As is clear from the spectrum fits (see especially the bottom right panel in Fig.~\ref{fig:korg_validation_giants}), the atomic data do not accurately capture the observations of the bluer of the two larger Fe lines within the spectrum (at $6646.93$~\AA); this behavior is essentially universal across the spectrum fits we have visually inspected. A small Fe line at $6645.37$~\AA~blended with the Eu feature is also not well-captured by the line list, and this becomes especially relevant at high metallicities (e.g., see especially Fig.~\ref{fig:knee} in Section~\ref{subsec:science_validation}). Presumably this issue might be addressed by updating the atomic data. However, given that we want to reproduce GALAH DR3 as closely as possible and use the metal-rich stars only as a check on our CNN, we find it most sensible to use to GALAH DR3 line list in its original form. From visual inspection, fits with \texttt{Korg} appear to be best for lower metallicity stars where the blended Fe line is not so prominent. Nonetheless, we find the relatively strong consistency between the \texttt{Korg} and GALAH DR3 abundances to indicate that this is a sufficiently reliable secondary check, regardless of occasional issues with the spectrum fitting.

In the Appendix (Section~\ref{subsec:korg_validation_dwarfs}), Fig.~\ref{fig:korg_validation_dwarfs} shows that the CNN's performance on dwarf stars passing our Eu line depth requirement is also excellent, with average absolute discrepancies in [Eu/H] between \texttt{Korg} and GALAH DR3 or the CNN both $<0.07$. Although the spread is slightly larger in the CNN validation data compared to \texttt{Korg} as compared to the dataset for GALAH DR3, we still note that $\sim75\%$ of the dwarfs in our CNN secondary validation subset have [Eu/H] residuals with \texttt{Korg} $<0.1$ and $\sim98\%$ have residuals $<0.2$. The standard deviation of the errors between \texttt{Korg} and the CNN is again low at a value of $0.052$. Given the difficulty of measuring such a weak line as Eu in dwarf stars, we consider this to be remarkably strong performance from the CNN.

\subsection{Metal-Poor Stars}
\label{subsec:metal_poor}

Metal-poor stars with $\mathrm{[Fe/H]}<-1$ are relatively rare in our dataset, comprising only $2500$ stars of the total sample. However, they are of broad community interest because they are predominantly halo stars \citep{Belokurov2022, Chandra_2024, Zhang_2024, Conroy_2022} and most Eu-enhanced stars are among this group \citep[e.g.,][]{Frebel_review_2023,RPA_DR1,RPA_DR5_2024}. Inconveniently, because of the natural tendency of neural networks to regress outliers towards the mean value in a dataset, we find that our CNN uniformly over-predicts [Eu/H] by a few $0.1$s of a dex at low metallicities (e.g., Fig.~\ref{fig:validation_1}). The predictions also become more uncertain with decreasing metallicity (e.g., Fig.~\ref{fig:best_practices_variance_params}, \ref{fig:all_stars_variance_params}). That the Eu line itself grows weaker in metal-poor stars likely further complicates matters.

Similarly to the high-[Eu/H] subset in \ref{subsec:loss}, we have tried upweighting metal-poor ([Fe/H] $< -1$) stars. However, while this upweighting led to lower \textit{reported} $\sigma_\textrm{[Eu/H]}$ for metal-poor stars in the validation dataset, it actually decreased the accuracy of their [Eu/H] predictions (while also significantly worsening the overall performance metrics). A similar decrease in prediction quality (increased random scatter) was observed in science validation tests for the DR4-only dataset, e.g., Fig.~\ref{fig:knee}. We have then attempted upsampling -- i.e. adding duplicated copies of metal-poor stars into the training/validation dataset to alleviate their underrepresentation -- noting a similar, though slightly lower, decrease in prediction quality. Finally, we have tried to lower the SNR requirements for the training/validation and prediction datasets to indirectly increase the representation of metal-poor stars. This also led to decreased prediction quality, including in particular, a significant increase in spurious outlier predictions.

In summary, despite several attempts to solve the problem, the network is not capable of generating high-precision predictions for the metal-poor subset -- possibly because the regions of attention within the metal-poor spectra simply do not contain enough signal. This effect is particularly visible in Fig.~\ref{fig:knee}, where we use [Eu/Fe] = [Eu/H]$_\text{pred}$ - [Fe/H]$_\text{DR4}$. For metal-poor stars the combination of (1) low [Fe/H] values, and (2) the network being unable to extract sufficient [Eu/H] information and instead regressing to the mean, generates an unrealistic excess of stars with very high ($\gtrsim 0.8$) [Eu/Fe].

Thus, because metal-poor stars are both complicated and also of community interest, we give them a more careful treatment here by providing a secondary catalogue of [Eu/H] abundances for all stars with [Fe/H]~$<-1$ from spectrum synthesis with \texttt{Korg}. Fitting with \texttt{Korg} is performed via the same procedures outlined for the secondary validation datasets in Section~\ref{subsec:korg}.
We emphasize that this catalogue is provided \emph{in addition} to the CNN catalogue, each with their respective benefits and drawbacks, and that users may choose whichever suits their scientific goals. A more careful discussion of the differences between the CNN and \texttt{Korg} can be found in the following section (Section~\ref{subsec:why_NN}), which also describes our choice not to use \texttt{Korg} for \emph{all} stars.

Notably, some stars in our \texttt{Korg} [Eu/H] catalogue have non-physical low abundances ($\mathrm{[Eu/H]}~\lesssim-4$) and correspondingly very large uncertainties on [Eu/H]. In these cases, \texttt{Korg} has simply been unable to detect an Eu line and has thus pushed the abundance to some very low value ([Eu/H]~$<-5$ or $-6$). We include a flag for these stars in the catalogue ($\texttt{lower\_limit\_flag}=1$ for affected stars) so users can easily remove them. Users may also chose to develop their own robust upper limits on the Eu abundances for these stars, if desired, although we do not choose to do so at this stage. In addition to these cuts, users should also apply the same criteria outlined in Section~\ref{subsec:data_cuts} for a clean selection of metal-poor \texttt{Korg} abundances with the exception of the \texttt{line\_depth} flag, which should not be used on this dataset as stars with Eu lines too weak to be measured by \texttt{Korg} will be filtered by the \texttt{lower\_limit\_flag}. Upon applying all of these cuts and the \texttt{lower\_limit\_flag}, there are $1\,589$ stars in our secondary metal-poor catalogue with $-1.7<\mathrm{[Fe/H]}<-1$, which form a metal-poor ``golden sample''.

\subsection{A Golden Sample of [Eu/H] Abundances}
\label{subsec:golden_sample}

We define a ``golden sample'' of high-confidence predictions as giants with $\texttt{eu\_flag}=0$; this sub-selection is also provided as a separate table for convenience. This is motivated by a systematic difference in precision of predictions between stellar types. Dwarfs constitute 48.5\% of the DR4-only dataset and their predicted [Eu/H] abundances are a valuable resource, which we make fully available. However, the quality of their [Eu/H] predictions is systematically lower due to strong selection effects on those inherently faint stars in the GALAH DR3 dataset.

The sample of dwarfs passing quality cuts is significantly more impacted by selection effects in GALAH DR3 than in GALAH DR4 (see e.g. Fig~\ref{fig:eu_line_strength}). The strong selection means that in the training dataset -- as compared to the prediction datasets -- dwarfs are underrepresented; furthermore, some ranges of stellar parameters and abundances are preferentially more represented for dwarfs. Therefore, [Eu/H] predictions for dwarfs suffer from strong interpolation or extrapolation and are burdened with higher uncertainties. We show this effect in more detail in Appendix \ref{sec:full_stellar_sample}. Nonetheless, as we note in Sec~\ref{subsec:korg}, CNN performance \emph{is} strong for dwarfs meeting our strict quality cuts; however, because these cuts include a selection for strong Eu lines, the dwarfs for which we have robust predictions are uniformly high in [Eu/H] and thus not reflective of the underlying distribution. For this reason, they are still excluded from the ``golden sample.''

\begin{figure}
	\includegraphics[width=\columnwidth, alt={The figure shows a series of 2D histograms with the distributions of the network's inferred standard deviations on the [Eu/H] prediction for the golden sample versus GALAH DR4 [Fe/H], [Ni/Fe], effective temperature, surface gravity, and the predicted [Eu/H] minus GALAH DR4 [Fe/H]. The distributions look very similar to those in Figure 8, with most of the standard deviations being between 0.05 and 0.1 and not having a strong variation relative to the variables presented. The overdensity at a surface gravity of 2.5, likely representing the red clump, is again present. The standard deviations also increase to slightly above 0.1 for stars with metallicities below -1 or hotter than about 5500 Kelvin, though there are relatively fewer of these stars.}]{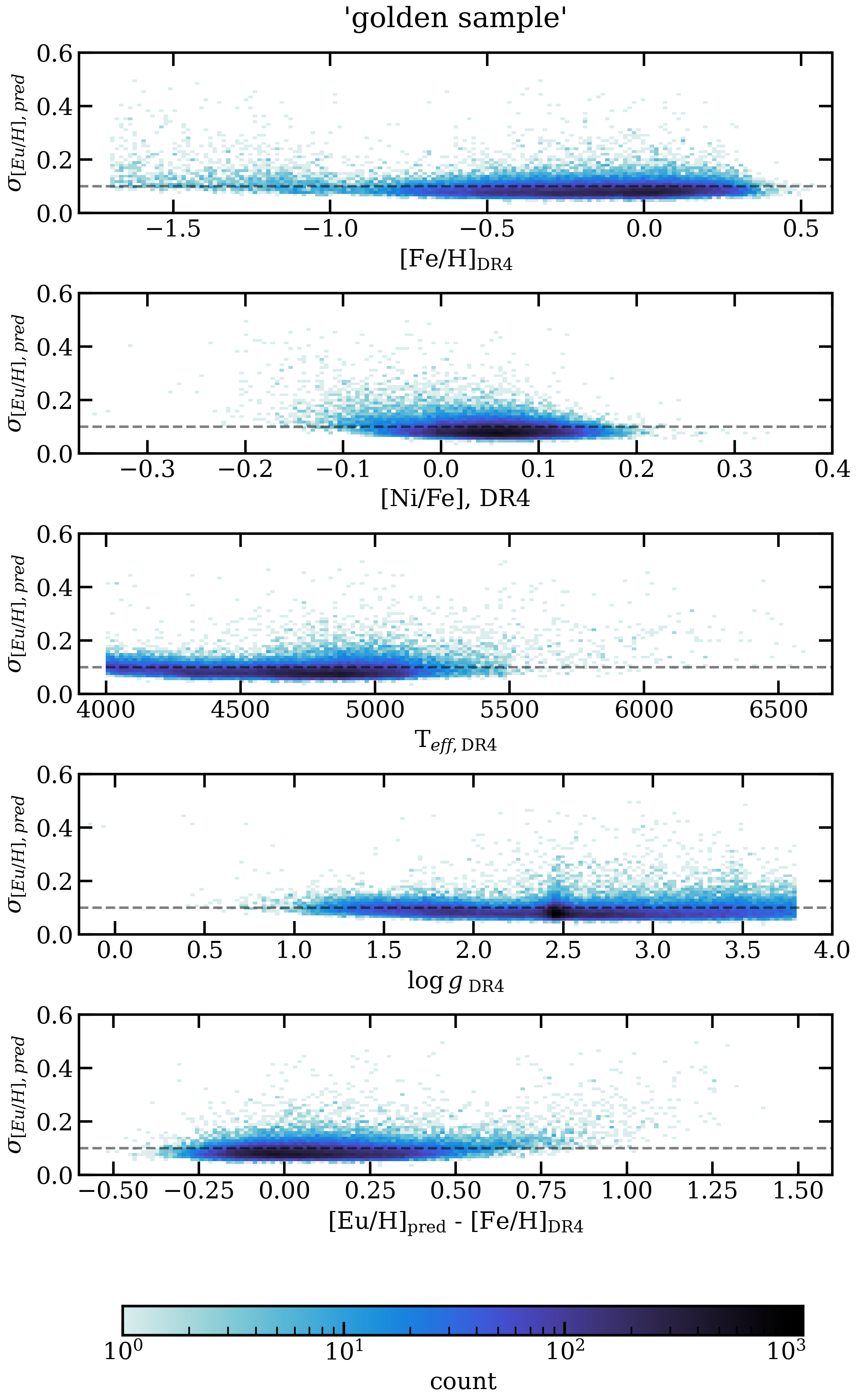}
   \caption{Like Fig.~\ref{fig:val_variance_params}, for the ``golden sample'' dataset. As [Eu/Fe]$_{\rm DR3}$ is not available for these stars, it is replaced by a predicted value of [Eu/H]$_{\text{pred}}$ - [Fe/H]$_{\text{DR4}}$ in the bottom plot. While the spread of $\sigma_{\text{[Eu/H], pred}}$ is larger than in Fig.~\ref{fig:val_variance_params}, the network still returns near-constant, $\lesssim$0.1 dex uncertainties for the bulk of [Eu/H] predictions within the ``golden sample'' dataset. CNN predictions for metal-poor stars are still included for illustrative purposes, although we suggest using the secondary \texttt{Korg} catalogue for these. (Note: 10 stars (0.02\% of the ``golden sample'' set) are outliers with $\sigma_{\text{[Eu/H], pred}}>0.6$ and were not included here for clarity of the figure.)}
    \label{fig:best_practices_variance_params}
\end{figure}

The ``golden sample'' comprises of $54\,068$ stars with CNN abundances for stars with [Fe/H]~$\geq-1$ and \texttt{Korg} abundances for those with $<-1.7$~[Fe/H]~$<-1$. We note that despite the restrictive cuts, they outnumber the crossmatch dataset from Section \ref{subsec:test_train}, i.e. all stars selected following GALAH DR3 and DR4 best practices with available DR3 Eu abundances. We visualize the uncertainties on their CNN label predictions in Fig.~\ref{fig:best_practices_variance_params}. We demonstrate that for the bulk of the ``golden sample'', we maintain the $\lesssim$0.1 dex precision of predictions. Users may apply additional cuts (e.g., on the network-generated [Eu/H] uncertainties) at their own discretion.

Notably, CNN predictions for our ``golden sample'' stars can be combined with the $52\,147$ DR4 stars that we use for training and validation (Section~\ref{subsec:test_train}), which have high-SNR spectra and robust DR3 [Eu/H] values. The golden sample is specifically designed to be representative of the training data from this sample, and the CNN produces predictions on the same scale as GALAH DR3; for complete consistency with the golden sample, one could choose to remove the dwarf stars from the DR3 cross-match. Thus, taken together, the golden sample and DR3 training and validation data constitute a homogeneous sample of $106\,215$ stars in GALAH DR4 with vetted europium abundances.

\section{Discussion}
\label{sec:discussion}

\subsection{A Note on Data-driven Abundances}
\label{subsec:why_NN}

\begin{figure*}
    \centering\includegraphics[width=1.7\columnwidth, alt={The clearest trend in the energy-angular momentum space is that the accreted stars, located at higher energies, have higher [Eu/Fe] and lower [Si/Fe] abundances than their in situ counterparts located at lower energies. The distinction is even more clear with [Eu/Si], with accreted stars again having much higher abundances. The dichotomy persists across all three metallicity groups, although the precise value of the abundances does vary with metallicity. The boundary between in situ and accreted stars is located at approximately -180000 km^2/s^2. We now give the approximate average [Si/Fe], [Eu/Fe], and [Eu/Si] values for in-situ and accreted stars in each of the metallicity bins. In the lowest metallicity bin where [Fe/H] is between -1.7 and -1.3, the in situ stars have [Si/Fe]=0.3, [Eu/Fe]=0.4, and [Eu/Si]=0.1. The accreted stars have [Si/Fe]=0.25, [Eu/Fe]=0.5, and [Eu/Si]=0.25. In the middle metallicity group with [Fe/H] between -1.3 and -1.9, the in situ stars have [Si/Fe]=0.35, [Eu/Fe]=0.4, and [Eu/Si]=0.1. The accreted stars have [Si/Fe]=0.2, [Eu/Fe]=0.5, and [Eu/Si]=0.3. In the highest metallicity group with [Fe/H] between -0.9 and -0.2, in situ stars have [Si/Fe] between 0.35, [Eu/Fe]=0.2, and [Eu/Si]=0.1. The accreted stars have [Si/Fe]=0.1, [Eu/Fe]=0.5, and [Eu/Si]=0.1. There are relatively few accreted stars in this lowest metallicity group, and the distinction between in situ and accreted stars is not as clear here in [Si/Fe] as it was in the more metal-poor group. In situ stars at higher energies and especially at higher z angular momenta have lower [Si/Fe] values between 0.1 and 0.25, while at lower energies [Si/Fe] is as high as approximately 0.35. Disk stars (at the highest angular momenta) also have the lowest [Eu/Fe] values in the highest metallicity group.}]{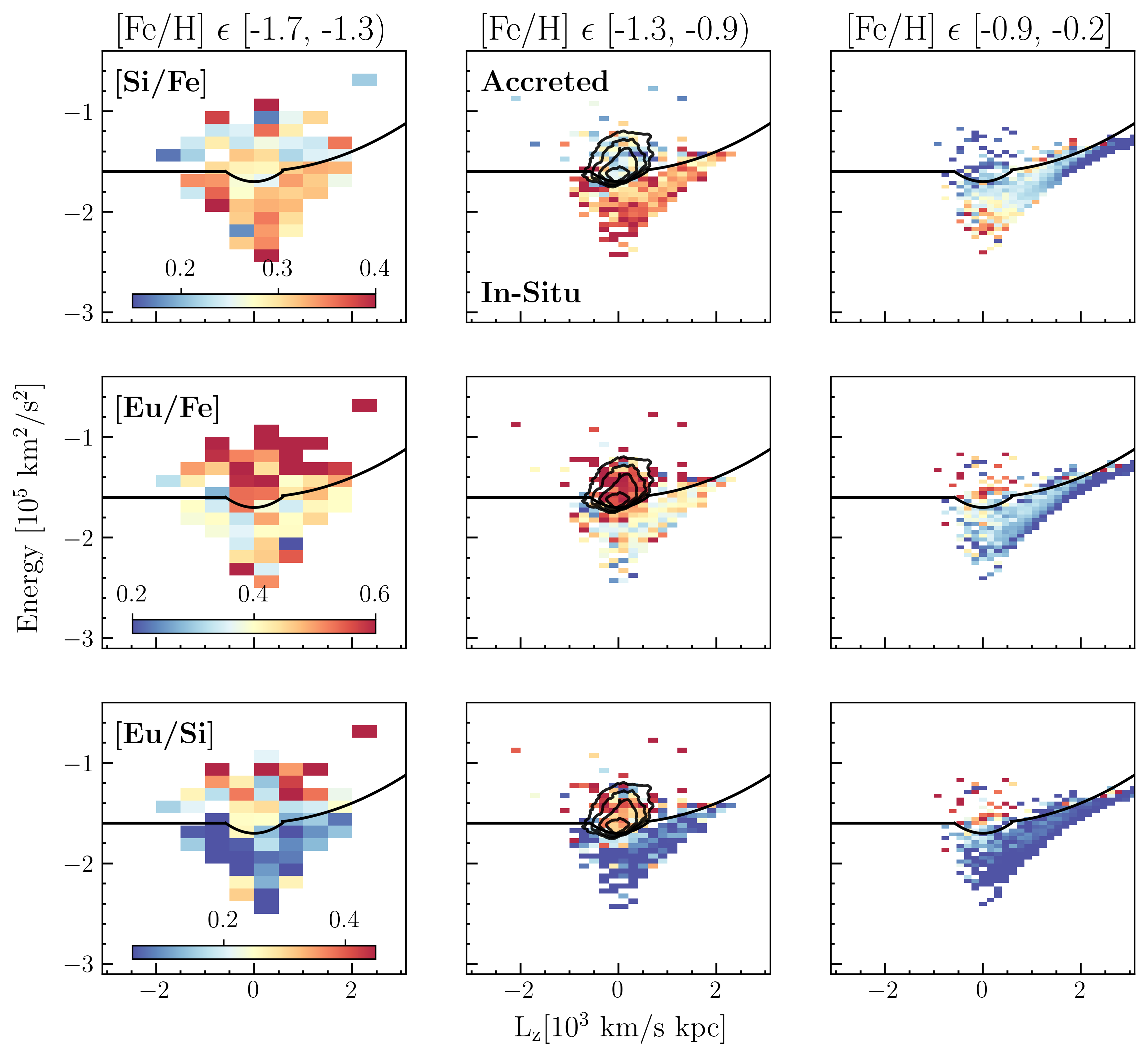}
    \caption{Created to be directly analogous to Fig.~1 of \citet{Monty_2024}, this figure shows the stars in our GALAH DR4 sample (see Sec.~\ref{subsec:data}) in the $E-L_z$ plane for the \citet{McMillan2017} potential. The solid black line indicates the separation between accreted and \textit{in situ} stars from \citet{Belokurov_Kravtsov_2023}, adapted for this potential. The stars are separated into metallicity bins of $-1.7\leq\mathrm{[Fe/H]}<-1.3$ (\textit{left}), $-1.3\leq\mathrm{[Fe/H]}<-0.9$ (\textit{middle}), and $-0.9\leq\mathrm{[Fe/H]}\leq-0.2$ (\textit{right}). The color-shading shows the mean [Eu/Fe] (\textit{top row}), [Si/Fe] (\textit{middle row}), [Eu/Si] (\textit{bottom row}) value per bin. The abundances are taken from our [Eu/H] predictions for this sample and the GALAH DR4 values of [Fe/H] and [Si/Fe] for each star. The black contours in the middle column are adapted from \citet{Belokurov_Kravtsov_2023} and mark the approximate location in the $E-L_z$ space of the debris from the Gaia-Sausage/Enceladus \citep[GS/E][]{Belokurov_GSE,Helmi_GSE} merger. Note that this figure almost exactly reproduces the style of the corresponding figure from \citet{Monty_2024}, with permission of the original authors.}
    \label{fig:Monty2024}
\end{figure*}

With the proliferation of machine learning-based approaches to the analysis of stellar spectra, it seems worthwhile here to examine our choice to primarily employ CNN-based abundances rather than more traditional techniques. Given our goal of reproducing GALAH DR3 [Eu/H] abundances from DR4, naturally the most faithful way to do this would be to redo the entire DR3 pipeline and get \emph{exact} DR3 abundances. However, aside from the actual work and time necessary for such an endeavor, doing so would necessarily mean regressing backwards from the changes implemented for DR4. One could no longer truly use the [Eu/H] values from this pipeline alongside the DR4 stellar parameters and thus the other abundances, but our goal is to provide a \emph{companion} Eu catalogue for the DR4 abundances themselves. A CNN is the logical choice for such a label transfer exercise. 

Our selection of stars -- high SNR, well-represented in the training set, and so forth -- is deliberately selected to provide the most robust and reliable abundance predictions possible. Nonetheless, we emphasize that these are \emph{predictions} and not ``measurements'' in the typical sense. Likewise, the CNN-predicted standard deviations on the [Eu/H] predictions, $\sigma_\textrm{[Eu/H],~pred}$ are not traditional measurement uncertainties but rather a measure of the variance in the network's prediction.

Given these considerations for our CNN-based [Eu/H] values, one could reasonably ask whether the Eu abundances could or should be redone with ``conventional" approaches (e.g., with \texttt{Korg}, which we used for validation in Section~\ref{subsec:korg}) in such a way that GALAH DR4's stellar parameters and other abundances can still be used. Although \texttt{Korg} provides more interpretable abundances than a neural network, we have two reasons to nonetheless estimate the bulk of our Eu abundances with the network and to use the synthesized abundances as a secondary check on the reliability of these network values. The first reason is a practical consideration: the network can estimate abundances far faster than \texttt{Korg} can synthesize them. For comparison, synthesizing abundances for the 200 stars from the secondary validation dataset took 1040.3~s; on the same system, predicting abundances with our model for a much larger sample of 10\,000 stars took 8.3~s with a single network and 166.6~s with a full ensemble. However, we recognize that other groups may have different computational resources at their disposal or smaller subsets of GALAH DR4 for which they want synthesized abundances, so we will also make the code with which we measured [Eu/H] with \texttt{Korg} public\footnote{The code is to be released on GitHub upon publication of this work.}. 

Second, our \texttt{Korg} [Eu/H] measurements are by their nature independent of the GALAH DR3 europium abundances, which makes them a valuable secondary comparison for our neural network results. Nonetheless, this independence means that the abundances we derive from \texttt{Korg} do not benefit from the extensive community testing and utilization undergone by the GALAH DR3 europium values. As is shown in Fig.~\ref{fig:korg_validation_giants}, there is a systematic offset between both GALAH DR3 and the CNN and \texttt{Korg}; the GALAH DR3 abundances are not precisely reproduced, nor could they be expected to be, using different stellar parameters and a different methodology. By contrast, the neural network [Eu/H] predictions are essentially the result of a simple label transfer exercise, with no different physical assumptions made via the use of a different radiative transfer code. Because the CNN's predictions are by design on a homogeneous scale with the GALAH DR3 [Eu/H] values, using the network allows us to combine our ``golden sample'' with DR3 abundances for a larger catalogue, as is suggested in Section~\ref{subsec:golden_sample}. Such a combination would not be so readily possible if we re-derived abundances with \texttt{Korg} or a different radiative transfer code. With the goal of extending the already broadly used GALAH DR3 europium measurements to a larger sample of stars, we thus provide the neural network predictions as our primary catalogue, with the necessary code to derive [Eu/H] from \texttt{Korg} included for secondary use as desired.

\subsection{Science Validation}
\label{subsec:science_validation}

As a final demonstration of the utility of our [Eu/H] abundances for GALAH DR4, we use our catalogue to reproduce known science results. These results use the ``golden sample'' of our DR4-exclusive giants (see Section~\ref{subsec:golden_sample}) and the corresponding catalogue of CNN abundances for stars with [Fe/H]~$\geq-1$ and the secondary catalogue of \texttt{Korg} abundances for stars with [Fe/H]~$<-1$. The goal of these final results is to provide additional testing and validation of our catalogue prior to community use. It is worth noting that although we present these recreations of known science results to demonstrate the utility of our europium abundances, we also used them iteratively to test and validate our [Eu/H] abundances throughout the process of developing the catalogue. These science tests often provided enlightening insight or helped to identify caveats in our method; for instance, it was from these science results that we first identified the need for cuts in [Fe/H] and $T_\mathrm{eff}$ (see Section~\ref{subsec:data_cuts} above). For this reason, we strongly recommend that those developing machine learning-based abundance catalogues undergo a similar process of science testing, as such tests can reveal nuances that are not immediately obvious from the results from a validation dataset (see Section~\ref{subsec:validation_performance}) alone.

\begin{figure}
    \includegraphics[width=0.97\columnwidth, alt={The figure shows three 2D histograms of the distributions of stars in the [Eu/Fe] versus [Fe/H] space, calculated with CNN Eu, a combination of Korg and CNN Eu, or GALAH DR3 Eu. [Fe/H] ranges between -2 and 0.5; the two new catalogs stop at -1.7 sharply but include somewhat more stars with [Fe/H] greater than 0 as compared to GALAH DR3. At [Fe/H] greater than -1, the CNN-only and CNN-Korg combination panels are identical as they both use the same values in this regime. [Eu/Fe] steadily decreases from a mean of approximately 0.5 at [Fe/H]=-1 to about -0.2 at [Fe/H]=0.2. The spread in [Eu/Fe] at any given [Fe/H] in this regime looks to be about 0.5. At [Fe/H] less than -1, the CNN abundances continue to increase with decreasing metallicity, up to an average [Eu/Fe] of about 0.7 at [Fe/H]=-1.7. By contrast, the Korg abundances for [Fe/H] less than -1 appear to have an approximately flat mean at [Eu/Fe]=0.5 and a spread of about 0.7. The GALAH DR3 [Eu/Fe] versus [Fe/H] trends look very similar to the CNN and Korg combined abundances, being approximately flat for [Fe/H] less than -1 and decreasing with increasing metallicity for [Fe/H] greater than -1. The difference is that there are a small number of stars at [Fe/H] greater than -1 with enhanced [Eu/Fe] as high as 0.5, which are not present in the CNN data. The Kobayashi et al (2020) model is approximately flat at a [Eu/Fe] of 0.45 for metallicities less than -1 and then begins to steadily decrease with increasing metallicity, reaching [Eu/Fe]=0 by [Fe/H]=0. The Zhao et al (2016) datapoints appear to follow a very similar trend, except there are only 4 points at [Fe/H] less than -1, and the data at [Fe/H]=0 extends as low as approximately [Eu/Fe]=-0.2. Both the Kobayashi et al (2020) model and the Zhao et al (2016) data appear approximately consistent with both GALAH DR3 and the combination of the CNN and Korg abundances; the CNN-only predictions do not match the flat trend for [Fe/H] less than -1.}]{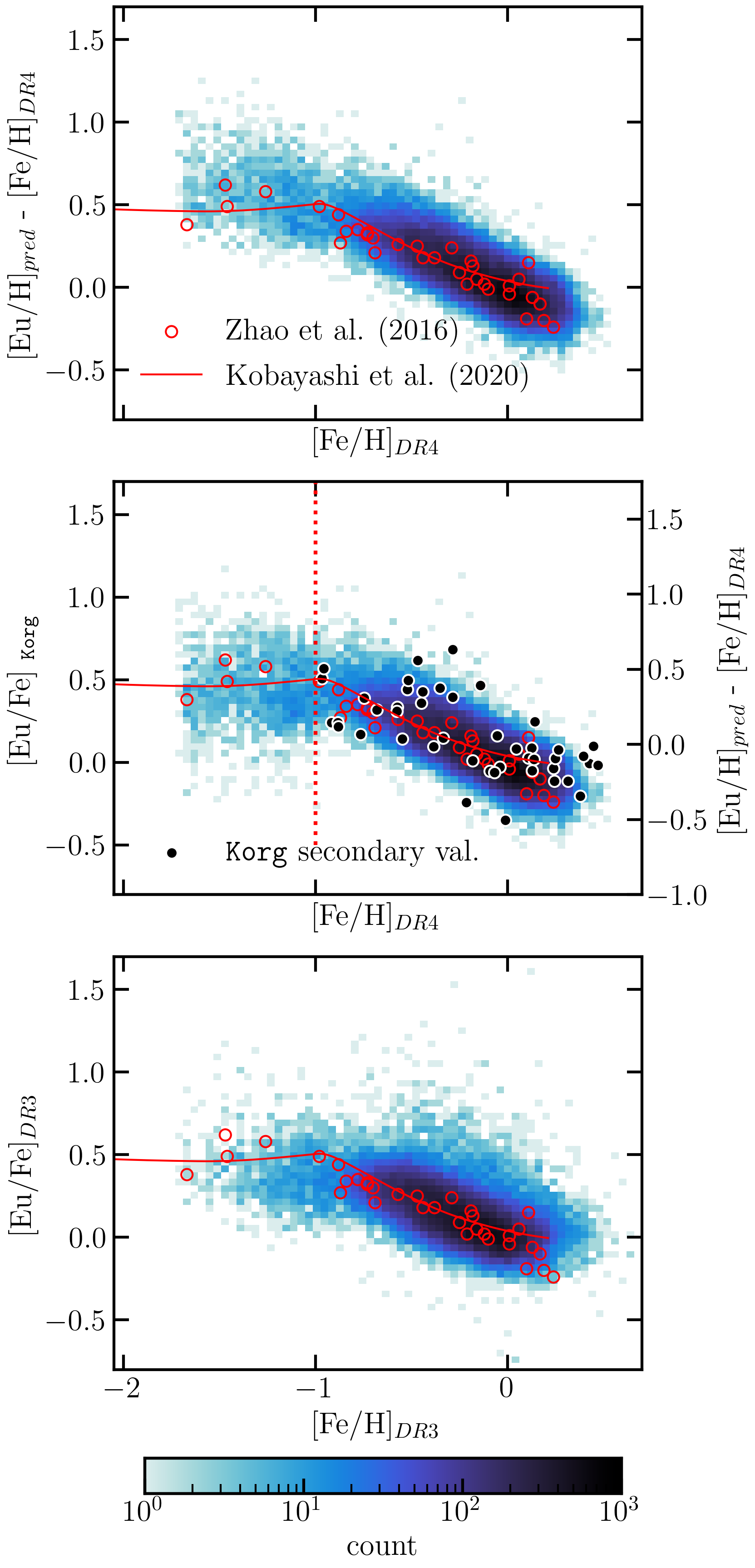}
    \caption{A comparison of [Fe/H] -- [Eu/Fe] trends in GALAH data extended with our catalogue predictions (blue 2D histogram) and in literature (red empty circles - datapoints from \citealt{Zhao2016}, red solid line - Solar neighborhood galactic chemical evolution model from \citealt{Kobayashi_2020_nucleosynthesis}). \textit{Top:} Using DR4 [Fe/H] and CNN-predicted [Eu/H] for all giants with $\texttt{eu\_flag}=0$. \textit{Center:} As top, but after replacing the y-axis positions for metal-poor stars ([Fe/H] $< -1$) with [Eu/Fe] synthetised with {\tt Korg} (our secondary catalogue; Section~\ref{subsec:metal_poor}); i.e., using the full ``golden sample''. The red vertical dotted line at [Fe/H] = -1 represents the cut. To demonstrate that reproducing the bend in the `knee' is not simply an effect of this cut, datapoints obtained with the same method from the \texttt{Korg} secondary validation dataset (Section~\ref{subsec:korg}) are overplotted as black circles for [Fe/H] $\geq-1$. \textit{Bottom:} As top, but using GALAH DR3 [Fe/H] and [Eu/Fe] for stars from the train/validation dataset (with a matching $X$-axis cut of [Fe/H] $>$ -1.7). The DR3 dataset shows very good agreement with \citet{Zhao2016,Kobayashi_2020_nucleosynthesis} (although with high scatter), but also strong underrepresentation of metal-poor stars compared to the DR4-only dataset.}
    \label{fig:knee}
\end{figure}

\citet{Monty_2024} used GALAH DR3 \citep{GALAH_DR3} to expand the selection of accreted stars via more conserved elemental abundances as done by \citet{Hawkins_2015, Das_2020, Buder_2022} with [Al/Fe] or [Na/Fe] vs. [Mg/Mn] in order to overcome the imperfect selection of stars via dynamical properties. \citet{Monty_2024} found that the ratio of [Eu/Si] differs between accreted and \textit{in-situ} stars; correspondingly, we show that this result can be reproduced with the GALAH DR4 [Si/Fe] and [Fe/H] values as well as our catalogue of CNN-derived DR4 [Eu/H] values. For consistency with that work, we adopt the same quality cuts for our data when possible, including $\texttt{flag\_si\_fe}=0$ and $\texttt{e\_si\_fe}<0.2$ for high quality DR4 [Si/Fe] measurements. The stellar parameter flag, $\texttt{flag\_sp}=0$, is included as part of our quality cuts in Section~\ref{subsec:data}. The SNR selection is necessarily not identical, as we impose a higher cut than the $\texttt{snr\_c3\_iraf}>30$ used in \citet{Monty_2024}, and likewise we do not use the [Fe/H] flag, per the GALAH DR4 recommended best practices.

In Fig.~\ref{fig:Monty2024}, we show results analogous to those from Fig.~1 of \citet{Monty_2024}\footnote{Two changes are made to Fig.~\ref{fig:Monty2024} as compared to Fig.~1 of \citet{Monty_2024}. First, the colorbar limits in the top panels are adjusted slightly to reflect the different [Si/Fe] values apparent in GALAH DR4 as compared to DR3; the [Eu/Fe] and [Si/Eu] colorbar limits are unchanged from \citet{Monty_2024} for consistency. Second, the lowest metallicity group is changed from a lower limit at $\textrm{[Fe/H]}=-2$ to $-1.7$, consistent with our recommended data cuts (Section~\ref{subsec:data_cuts}).}, with mean stellar [Si/Fe], [Eu/Fe], and [Eu/Si] ratios shown in the $E-L_\mathrm{z}$ plane from the \citet{McMillan2017} MW potential (as used in the GALAH DR3 and DR4 dynamics value-added catalogues). The division between regions dominated by accreted and \textit{in-situ} stars, marked in black in each panel, is adopted from \citet{Belokurov_Kravtsov_2023} and marks a clear difference in Eu and Si abundances. In particular, accreted (high energy) stars tend to be lower in [Si/Fe] and higher in [Eu/Fe] than their \textit{in-situ} (low energy) counterparts. Correspondingly, the accreted stars also have a higher [Eu/Si] abundance, with this distinction being the most clear in the middle metallicity ($-1.3\leq\mathrm{[Fe/H]}<-0.9$) range. The highest metallicity group ($-0.9\leq\mathrm{[Fe/H]}\leq-0.2$) has fewer accreted stars with which to draw a distinction, although there is some hint that the difference in [Eu/Si] ratio remains. The accreted and \textit{in-situ} stars in the $-1.7\leq\mathrm{[Fe/H]}<-1.3$ group show less contrast in their [Eu/Si] abundances than the higher metallicity group, although some difference remains across the boundary; this result was also found by \citet{Monty_2024} using GALAH DR3. Our catalogue thus provides a promising avenue to select more accreted stars via their [Eu/Si] abundances, including in the regions with lowest orbit energies that are dominated by in-situ stars in dynamical selection (see e.g., Fig.~9d in \citealt{Buder_2022} and Fig.~5 in \citealt{Buder_2025}). This finding is also consistent with the work of \citet{Matsuno_2021,Aguado_2021}, who find higher \textit{r}-process abundances among accreted stars.

As another validation test, we also show that our new catalogue of Eu abundances reproduce known trends with [Fe/H]. As noted by \citet{Monty_2024}, these patterns are impossible to interpret as a single chemical enrichment history due to the contribution of both the \textit{in situ} Milky Way itself and various accreted dwarf galaxies, but our goal here is merely to demonstrate that our abundances match the overall patterns found in other literature sources rather than to sketch a consistent enrichment history. In Fig.~\ref{fig:knee}, we show [Eu/Fe] versus [Fe/H] trends using our CNN predictions, a combination of our CNN predictions and the \texttt{Korg} metal-poor star secondary catalogue, and DR3 values, as a comparison. Beginning with the metal-poor stars with [Fe/H]~$\lesssim-1$, the CNN-based predictions for [Eu/Fe] are uniformly high in this range and steadily increase with decreasing metallicity. This trend is not fully in agreement to that seen in the GALAH DR3 data as well as the comparison set of Solar neighborhood dwarfs from \citet{Zhao2016} (who measured Eu predominantly from the bluer line at $4129.72$~\AA). As we discuss in Section~\ref{subsec:metal_poor}, we believe this behaviour is non-physical and caused mostly by regression towards the mean, thus well-motivating the use of our secondary \texttt{Korg} catalogue of abundances in this regime. (Interestingly, if we examine DR3 [Eu/H] values with DR4 metallicities such that $\textrm{[Eu/Fe]}=\textrm{[Eu/H]}_\textrm{DR3}-\textrm{[Fe/H]}_\textrm{DR4}$, the ``knee'' also becomes less pronounced. This shows that the network is not significantly deviating from DR3 [Eu/H] patterns, and that in addition to previously discussed issues the metal-poor sample also experiences strong DR3-DR4 offsets.) Looking at the \texttt{Korg} abundances for metal-poor stars in the middle panel of Fig.~\ref{fig:knee}, we note a visibly different behavior, wherein the [Eu/Fe] abundances are approximately stable relative to metallicity but exhibit a wide spread to low and high values. This is consistent with other literature work which find the halo to show a broad variation in Eu abundances \citep[e.g.,][]{Cote_2019} and also appears approximately similar to the behavior seen among the GALAH DR3 stars (see bottom panel of Fig.~\ref{fig:knee}). The extent to high [Eu/Fe] values in this low metallicity regime suggests exciting prospects for our catalogue to identify new \textit{r}-process enhanced stars in the halo.

At a metallicity of approximately [Fe/H]~$\approx-0.9$ in the the \texttt{Korg} and CNN combined dataset, there is a ``knee'' feature wherein the [Eu/Fe] values shift from a flat average with a wide variation at low metallicities to more uniform, lower abundances at high metallicities. This ``knee'' is also apparent in the GALAH DR3 data and in the galactic chemical evolution model from \citet{Kobayashi_2020_nucleosynthesis}, where it is caused by the increased Fe production from the onset of Type Ia supernovae (SNe), although it is not so readily visible in the CNN-only data (top panel of Fig.~\ref{fig:knee}) due to the consistent over-prediction of [Eu/H] for metal-poor stars.
For higher metallicities, ([Fe/H]~$\gtrsim-0.9$), the CNN-predicted europium abundances steadily decrease with increasing metallicity \citep[consistent with][among many others]{Zhao2016,Battistini_Bensby_2016}. This behavior is also in agreement with that of GALAH DR3 and the galactic chemical evolution model from \citet{Kobayashi_2020_nucleosynthesis}.
Notably, for [Fe/H]~$>-1$, the CNN-predicted [Eu/Fe] values do show somewhat less spread than those from GALAH DR3, especially to higher values in [Eu/Fe]. Inspection of the GALAH DR3 data shows that most of these high-[Eu/Fe] stars are dwarfs, which, as discussed in Section~\ref{subsec:golden_sample}, are skewed towards high Eu abundances due to the necessity of having strong lines to measure and the inherent difficulties involved in working with the dwarf spectra. Because the ``golden sample'' we use for DR4 includes only giants, the corresponding dwarf stars to those in DR3 are not included.
We have found that the CNN \emph{is} able to recover interesting, anomalously Eu-enhanced giants in the Galactic disk. We leave the exploration of these unusual and exciting stars for future work (Kane et al. in prep.).

Given that the ``knee'' feature occurs close to [Fe/H]~$\approx-1$, the same metallicity at which we switch from \texttt{Korg} to CNN abundances, we seek to validate that this feature is real and not an artifact of the joining of the two datasets. It is reassuring that this turnover feature appears both in the GALAH DR3 data and also repeatedly in literature studies, but we further test it by deriving a set of Eu abundances for $50$ stars with [Fe/H]~$>-1$ with \texttt{Korg} to confirm that the knee persists within the synthesized abundances alone. As is clear from this additional dataset visualized in the middle panel of Fig.~\ref{fig:knee}, the turnover \emph{does} in fact persist in the \texttt{Korg}-exclusive data marked as black circles, and this combined with the fact that the same feature also exists in multiple other sources reassures that it is physical and not a product of a CNN-\texttt{Korg} offset. In fact, contrary to considering this feature a product of the joining of these two catalogues, it may be the case that the change in behavior in [Eu/Fe] in this regime is why the CNN performance declined at low metallicities, thus necessitating the use of the secondary \texttt{Korg} data.
At metallicities [Fe/H]~$\gtrsim0$, the \texttt{Korg} Eu abundances clearly diverge to higher values than those exhibited by either the CNN or GALAH DR3; as is discussed briefly in Section~\ref{subsec:korg}, visual inspection of the spectra shows this to be caused by poor fits due to incomplete accounting for the Fe line blended with the Eu feature, which is most prominent at the highest metallicities. However, because the \texttt{Korg} abundances here are intended merely as a check on the turnover feature, which occurs at much lower metallicities where the spectral synthesis provides better fits, we do not further address this discrepancy here.

In conclusion, we are satisfied with the behaviour of our CNN-predicted Eu abundances and confident that the golden sample has been well-vetted to remove spurious data and/or unphysical results.

\section{Conclusions}
\label{sec:conclusions}

We have presented in this work a new catalogue of [Eu/H] values inferred via a convolutional neural network (CNN) to complement the catalogue of GALAH DR4. In particular, we infer europium abundances observations by training the neural network to predict GALAH DR3 [Eu/H] values and inferred uncertainties from the DR4 stellar parameters and high SNR ($>50$) spectra. In validation testing, the network shows good performance across a range of stellar parameters, with the average residuals between GALAH DR3's and the CNN's inferred [Eu/H] values to be $\lesssim0.1$ in essentially all regions of the Kiel diagram. Secondary testing with the \texttt{Korg} \citep{KORG_2023,Korg_2024} LTE synthesis code also confirms [Eu/H] abundances within an average of $\sim0.1$ of the CNN predictions for both giant and dwarf stars.

\begin{enumerate}
    \item To produce the most robust catalogue possible, in Section~\ref{subsec:data_cuts} we provide a series of quality cuts for our inferred abundances. The cuts involve flags for stellar parameter ranges in which our predictions lack reliability ($T_\mathrm{eff}<4000$~K and [Fe/H]~$<-1.7$), spectra where the continuum flux diverges from $1$, and a flag for Eu line depth, wherein only stars with strong lines are retained in the catalogue to reflect the properties training data.
    \item Our suggested cuts yield a ``golden sample'' of $54\,068$ giant stars as well as $3\,630$ strong-line dwarfs, which is described in detail in Section~\ref{subsec:golden_sample}. Note that our ``golden sample'' alone still exceeds the number of high-SNR spectra in our DR3 training and validation sets, which contain $\sim52\,000$ stars.
    \item The one regime where our CNN shows markedly worse performance is for metal-poor stars with [Fe/H]~$<-1$, where [Eu/H] values are systematically over-predicted. We believe that this effect arises because metal-poor stars are less common in the training data than stars with [Fe/H]~$>-1$ and because they have lower [Eu/H] abundances than their metal-rich counterparts. A tendency to ``regress towards the mean,'' which is a typical behavior for neural networks, thus pushes these values towards systematic overestimates on the order of a few $0.1$s of a dex.
    \item After attempting several strategies with the neural network to mitigate this behavior, we found that the best course of action was to provide a secondary catalogue of Eu abundances derived using \texttt{Korg} for all stars with [Fe/H]~$<-1$. Although we recognize that this is an imperfect solution, we nonetheless recommend this secondary catalogue for use for metal-poor stars and find that it generally well reproduces the overall behavior of the DR3 europium abundances. A complete discussion of the metal poor stars in our catalogue can be found in Section~\ref{subsec:metal_poor}.
    \item In addition to validation testing with DR3 abundances and synthesis via \texttt{Korg}, we demonstrate in Section~\ref{subsec:science_validation} that our new catalogue of abundances is capable of reproducing known science results. These results are produced with our ``golden sample'' of high-confidence giant stars. First, following from \citet{Monty_2024}, we demonstrate that the GALAH DR4 [Eu/Si] abundances using our CNN-derived [Eu/H] differ between accreted and \textit{in situ} MW stars. This result persists across a broad metallicity range of $-1.7<\textrm{[Fe/H}]<-0.2$, using the \texttt{Korg}-derived abundances for stars with [Fe/H]~$<-1$. We also reproduce the known patterns of [Eu/Fe] versus [Fe/H]. Below [Fe/H]~$\approx-1$, our \texttt{Korg} Eu abundances exhibit substantial scatter, reflecting the stochastic nature of \textit{r}-process production; this is also the regime in which most \textit{r}-process enhanced stars appear. At [Fe/H]~$\approx-1$, there is an inflection point or ``knee'' representing the onset of Type Ia SNe and well-captured by galactic chemical evolution models \citep{Kobayashi_2020_nucleosynthesis}, and beyond this value the CNN-based [Eu/Fe] abundances steadily decrease with increasing metallicity and are far more homogeneous than those found among metal-poor stars.
\end{enumerate}

With these tests, we hope this catalogue proves useful to the community as an expansion of the already-available GALAH europium abundances from previous data releases and as a companion catalogue to the most recent DR4. Included are well over a $100\,000$ new [Eu/H] predictions, including over $54\,000$ abundances for giant stars with especially high precision. When our ``golden sample'' is taken in combination with the $\sim52\,000$ GALAH DR3 [Eu/H] values we use for training and validation, there are now $106\,215$ GALAH DR4 stars with high-SNR observations which have Eu abundances available, making this one of the largest catalogues of stellar europium abundances in the literature. The possibilities for such a catalogue are numerous, including identifying new \textit{r}-process enhanced stars throughout the Milky Way, modeling the Galactic chemical evolution due to rare nucleosynthetic events \citep[e.g.,][]{Kobayashi_2020_nucleosynthesis}, and selecting accreted stars via their [Eu/$\alpha$] abundances (as in \citealt{Matsuno_2021, Monty_2024}; see also \citealt{Aguado_2021}).
More developments in our knowledge of the stellar europium content in the Galaxy are expected in the coming years with the 4-metre Multi-Object Spectroscopic Telescope \citep[4MOST,][]{4MOST_deJong_2019,4MOST_Walcher_2019} survey, which will also use data-driven abundance derivation techniques via the Payne \citep{Payne_YSTing_2019}. Our catalogue thus represents a new step forward in two separate but related topics of broad community interest: machine learning-based stellar abundance determination and the exploration of \textit{r}-process abundances across the Galaxy.

\section*{Acknowledgements}

The authors would like to thank the organizers and lecturers at the Galaxy Modeling Workshop at the University of Surrey from December, 2024. This work originated as a project at that workshop and would not have been possible without its organization. The members of the SOC/LOC were Natsuki Funakoshi, Jason Hunt, and Daisuke Kawata. The lecturers for the workshop were Payel Das, Denis Erkal, Stephanie Monty, Mike Petersen, Jason Sanders, and Eugene Vasiliev. The authors would also like to thank Adam Wheeler, Vasily Belokurov, Joachim Wambsganss, Tadafumi Matsuno, Carrie Filion, GyuChul Myeong, Madeleine McKenzie, Hanyuan Zhang, David Chemaly, and Tom Hehir for useful conversations in the development of this work.

SGK acknowledges PhD funding from the Marshall Scholarship, which is jointly supported by the UK government, the Cambridge Trust, and Trinity College, Cambridge. 
ZK is a Fellow of the International Max Planck Research School for Astronomy and Cosmic Physics at the University of Heidelberg (IMPRS-HD).
SB acknowledges support from the Australian Research Council under grant number DE240100150.

This work made use of the Fourth Data Release of the GALAH Survey (Buder et al. 2021). The GALAH Survey is based on data acquired through the Australian Astronomical Observatory, under programs: A/2013B/13 (The GALAH pilot survey); A/2014A/25, A/2015A/19, A2017A/18 (The GALAH survey phase 1); A2018A/18 (Open clusters with HERMES); A2019A/1 (Hierarchical star formation in Ori OB1); A2019A/15, A/2020B/23, R/2022B/5, R/2023A/4, R2023B/5 (The GALAH survey phase 2); A/2015B/19, A/2016A/22, A/2016B/10, A/2017B/16, A/2018B/15 (The HERMES-TESS program); A/2015A/3, A/2015B/1, A/2015B/19, A/2016A/22, A/2016B/12, A/2017A/14, A/2020B/14 (The HERMES K2-follow-up program); R/2022B/02 and A/2023A/09 (Combining asteroseismology and spectroscopy in K2); A/2023A/8 (Resolving the chemical fingerprints of Milky Way mergers); and A/2023B/4 (s-process variations in southern globular clusters). We acknowledge the traditional owners of the land on which the AAT stands, the Gamilaraay people, and pay our respects to elders past and present. This paper includes data that has been provided by AAO Data Central (datacentral.org.au).

This work has made use of data from the European Space Agency (ESA) mission
{\it Gaia} (\url{https://www.cosmos.esa.int/gaia}), processed by the {\it Gaia}
Data Processing and Analysis Consortium (DPAC,
\url{https://www.cosmos.esa.int/web/gaia/dpac/consortium}). Funding for the DPAC
has been provided by national institutions, in particular the institutions
participating in the {\it Gaia} Multilateral Agreement.

This research has made use of NASA's Astrophysics Data System Bibliographic Services, \texttt{NumPy} \citep{Harris2020}, \texttt{SciPy} \citep{Virtanen2020}, \texttt{Matplotlib} \citep{Hunter2007}, \texttt{Astropy} \citep{astropy:2013,astropy:2018,astropy:2022}, \texttt{scikit-learn} \citep{scikit-learn}, {\tt pandas} \citep{reback2020pandas, mckinney-proc-scipy-2010}, {\tt PyTorch} \citep{Paszke_pytorch}, {\tt SciencePlots} \citep{scienceplots}, {\tt CMasher} \citep{cmasher}, and \texttt{Korg} \citep{KORG_2023,Korg_2024}. We acknowledge the use of Paul Tol's colourblind-friendly colour palettes\footnote{\url{https://sronpersonalpages.nl/~pault/}}.

\section*{Data Availability}

 This work relies on publicly available data from GALAH Data Release 3 and 4 and from the \textit{Gaia} mission. The dataset of new [Eu/H] abundances generated in this work is publicly available on \href{https://doi.org/10.5281/zenodo.17737173}{Zenodo}. Also available in the Zenodo folder is a file of the interpolated GALAH spectra used as input for the neural network (see Section~\ref{subsec:prediction_set}). This dataset will also be made available on DataCentral\footnote{\url{https://cloud.datacentral.org.au/teamdata/GALAH/public/GALAH_DR4/catalogs/galah_dr4_vac_europium/}} as a value-added catalogue alongside GALAH DR4\footnote{\url{https://www.galah-survey.org/dr4/the_catalogues/}}. The code underlying this work will be made public upon publication of the paper.



\bibliographystyle{mnras}
\bibliography{refs} 




\appendix
\section{Contents of the [Eu/H] Catalogues}
\label{sec:catalogue_contents}

We provide three catalogues to users, which are as follows: 1.) the catalogue of CNN-derived abundances for DR4-exclusive stars, 2.) the secondary catalogue of \texttt{Korg}-derived abundances for DR4-exclusive stars with [Fe/H]~$<-1$, and 3.) the ``golden sample'' catalogue (Sec.~\ref{subsec:golden_sample}), which combines catalogues (1) and (2) for giant stars meeting the relevant quality cuts. Samples of the main and secondary catalogues are presented in Tables~\ref{tab:main_catalogue} and~\ref{tab:secondary_metal_poor}, respectively; all three datasets are available in the online supplementary material.

\begin{table*}
\centering
\begin{tabular}{lccccccccccc} 
\hline
\hline
\texttt{sobject\_id} & $T_{\rm eff}$ [K] & [Fe/H] & $\log g$ & [Ni/Fe] & [Eu/H]$_{\rm pred}$ & $\sigma_{\rm [Eu/H], pred}$ & \texttt{line\_depth} & flags \\
\hline
131216001101002 & 4932 & -0.417 & 3.124 & 0.075 & -0.017 & 0.119 & 0.020 & (0, 0, 0, 0) \\
131216002101285 & 4987 & -0.102 & 4.361 & -0.034 & 0.359 & 0.246 & 0.002 & (0, 0, 1, 1) \\
131217001801084 & 4937 & 0.323 & 4.620 & 0.072 & 0.621 & 0.205 & 0.002 & (0, 0, 1, 1) \\
131217002301114 & 5083 & 0.107 & 4.063 & 0.027 & 0.306 & 0.090 & 0.017 & (0, 0, 1, 1) \\
131217002801111 & 4897 & -0.039 & 4.391 & 0.006 & 0.451 & 0.137 & 0.020 & (0, 0, 0, 0) \\
... & ... & ... & ... & ... & ... & ... & ... & ... \\

 \hline
 
\end{tabular}
\caption{Sample table containing the first 5 rows of the main catalogue, including stellar parameters, CNN predictions and quality cuts for GALAH DR4 stars. The first column lists the unique GALAH DR4 \texttt{sobject\_id} identifiers. The publicly available online dataset also lists 2MASS (\texttt{tmass\_id}) and \textit{Gaia} DR3 (\texttt{gaiadr3\_source\_id}) IDs. Flags listed in the last column are included in separate columns in the supplementary dataset as \texttt{stellar\_params\_flag}, \texttt{continuum\_flux\_flag}, \texttt{line\_depth\_flag} and \texttt{eu\_flag}, respectively. The full table containing 118 946 stars is available in the online supplementary material.}
\label{tab:main_catalogue}
\end{table*}

\begin{table*}
\centering
\begin{tabular}{lccccccccccc} 
\hline
\hline
\texttt{sobject\_id} & $T_{\rm eff}$ [K] & [Fe/H] & $\log g$ & [Ni/Fe] & [Eu/H]$_{\tt Korg}$ & $\sigma_{\rm [Eu/H], \texttt{Korg}}$ & flags \\
\hline
140118003501046 & 3849 & -1.053 & 4.409 & -0.070 & -5.182 & 45854.438 &  (1, 0, 1, 1)\\
140305003201024 & 5054 & -2.021 & 1.937 & -0.147 & -1.477 & 0.048 &  (1, 0, 0, 1)\\
140307003101118 & 4739 & -1.840 & 1.678 & -0.105 & -1.856 & 0.072 &  (1, 0, 0, 1)\\
140307003101219 & 4880 & -1.692 & 1.816 & -0.135 & -5.715 & 59377.358 &  (0, 0, 1, 1)\\
140308000101112 & 5128 & -1.063 & 3.087 & 0.027 & -0.813 & 0.005 &  (0, 0, 0, 0)\\
... & ... & ... & ... & ... & ... & ... & ...  \\

 \hline
 
\end{tabular}
\caption{Sample table containing the first 5 rows of the secondary catalogue for metal-poor stars. The first five columns are defined as in Table~\ref{tab:main_catalogue}, followed by [Eu/H] values and uncertainties synthetised with {\tt Korg}. The \texttt{line\_depth\_flag} is replaced by \texttt{lower\_limit\_flag} to reflect a different treatment of weak-line stars with this method. The full table containing 2500 stars is available in the online supplementary material.}
\label{tab:secondary_metal_poor}
\end{table*}

\section{Secondary \texttt{Korg} validation for dwarf stars}
\label{subsec:korg_validation_dwarfs}

\begin{figure*}
    \includegraphics[width=2\columnwidth, alt={The left two panels show scatter plots of the GALAH DR3 or CNN predicted [Eu/H] versus the Korg-derived [Eu/H], respectively, with a 1-to-1 line overplotted for reference. The mean absolute difference between GALAH DR3 and Korg is 0.0627, and between the CNN and Korg it is 0.0696. [Eu/H] values in both panels range between approximately -1.1 and 0.6. There is some substantial scatter on the order of 0.05 and 0.1 dex between Korg and GALAH DR3 or the CNN, and this scatter appears slightly larger for the CNN abundances. The typical CNN-inferred uncertainty in [Eu/H] is about 0.1 and about 0.07 from Korg. The right panel shows three example GALAH spectra in our wavelength region and their associated fits with Korg. The spectra generally have good Korg fits to the Eu line except for the Fe line at 6645.37 angstroms which is visibly blended with the Eu line and is not matched by Kor in any of the three spectra. The rest of the spectrum regions, especially the continua, are well fit by Korg with the exception of the iron line at 6646.93 angstroms, which is again not fit in any of the three spectra and only appears to dip slightly below the continuum level in the Korg fit.}]{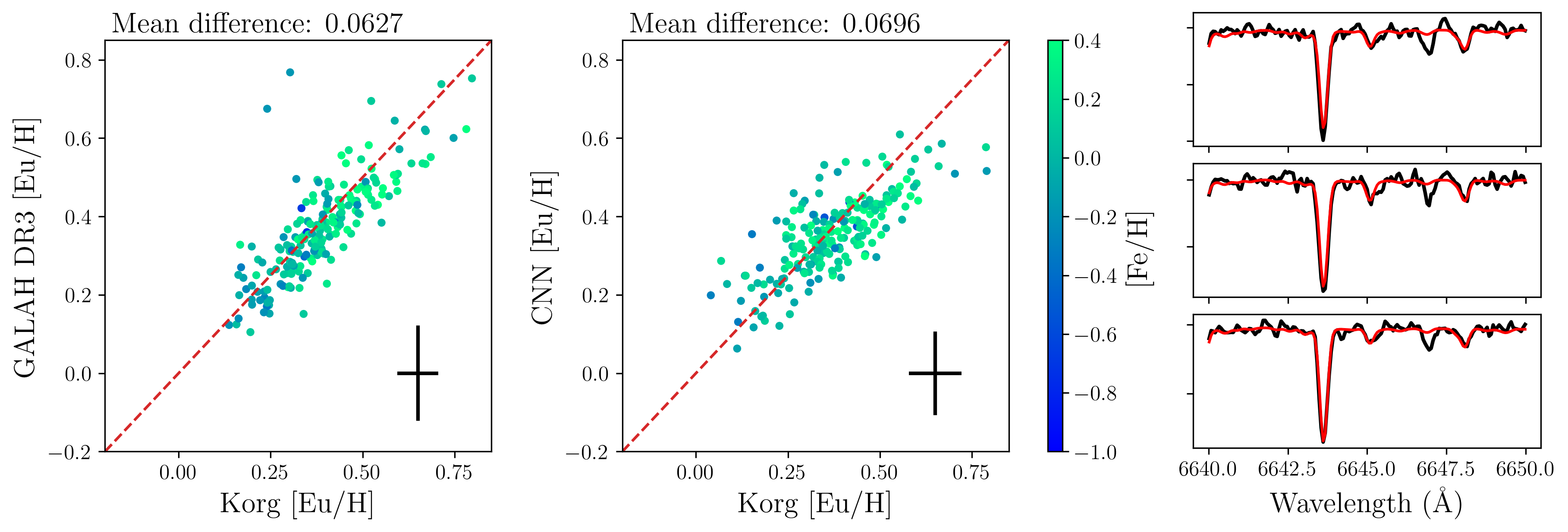}
    \caption{As in Fig.~\ref{fig:korg_validation_giants}, the \texttt{Korg} comparison of [Eu/H] abundances, now for a set of $400$ dwarfs which pass our quality cuts, including the requirement for Eu line depth. As before, the left panel shows a comparison of GALAH DR3 and \texttt{Korg}-derived [Eu/H] values, while the middle panel shows [Eu/H] from the CNN versus \texttt{Korg}. Both are again color-coded by the stellar [Fe/H]. The error markers again show the average uncertainties. The mean absolute difference between the two sets of [Eu/H] values is also shown for both panels. The right three panels show three sample spectra from the DR4-exclusive dataset in our selected wavelength region. The black lines mark the observed GALAH spectrum, while the red line is the best fit flux from \texttt{Korg}. Although the axis limits span a smaller range than those in Fig.~\ref{fig:korg_validation_giants} for visualization purposes, note that the mean absolute differences between the \texttt{Korg} [Eu/H] values and those from either GALAH DR3 or the CNN are actually slightly smaller for the dwarfs than the giants.}
    \label{fig:korg_validation_dwarfs}
\end{figure*}

Fig.~\ref{fig:korg_validation_dwarfs} shows our secondary validation with \texttt{Korg} for dwarf stars passing our quality cuts (see Section~\ref{subsec:data_cuts}). An associated discussion of this figure can be found in Section~\ref{subsec:korg}.

\section{Prediction uncertainties for the Full Stellar Sample}
\label{sec:full_stellar_sample}

\begin{figure*}
 \centering
	\includegraphics[width=\textwidth, alt={The figure shows panels of 2D histograms of the distributions of the CNN-inferred standard deviation on the [Eu/H] prediction versus GALAH DR4 [Fe/H], [Ni/Fe], effective temperature, surface gravity, and predicted [Eu/H] - GALAH DR4 [Fe/H]. These distributions are shown separately for dwarf and giant stars. For the giants, many of the standard deviations appear to fall around 0.1, but there is a much larger spread to higher values than there was in the golden sample (see Fig. 13), with some standard deviations as high as 0.5. There is a particularly sharp increase in the standard deviations for giants at effective temperatures below 4000 Kelvin, although higher standard deviations also appear at metallicities below -2, surface gravities below 1, and predicted [Eu/H] minus GALAH DR4 [Fe/H] below -0.2. For the dwarf stars, the standard deviations span again from approximately 0.1 to 0.5, but they are much less concentrated at 0.1 and thus there are more dwarf stars with higher uncertainties. The uncertainties are especially high for hot dwarfs with effective temperatures greater than 6000 K.}]{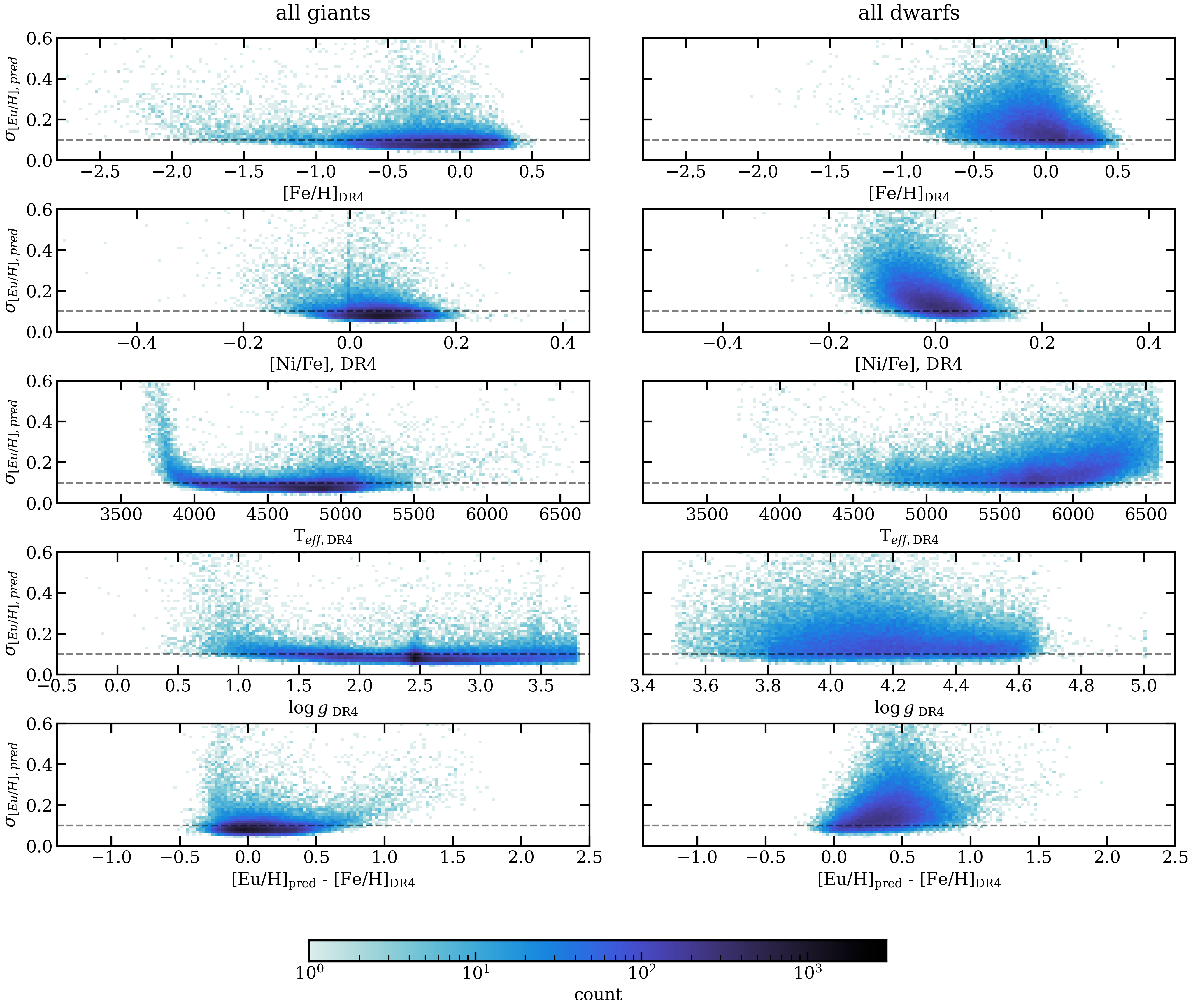}
   \caption{Like Fig.~\ref{fig:best_practices_variance_params}, but for all stars in the catalogue regardless of flags and quality cuts. The dataset is split into giants (\textit{left}) and dwarfs (\textit{right}); the subsets contain 61215 and 57731 stars, respectively. 1021 giants (1.7\% of all giants) and 395 dwarfs (0.7\% of all dwarfs) are outliers with $\sigma_{\text{[Eu/H], pred}}>0.6$ and were not included here for clarity of the figure; we discuss those outliers in detail in the Section~\ref{sec:full_stellar_sample} text.}
   \label{fig:all_stars_variance_params}
\end{figure*}

In Fig.~\ref{fig:all_stars_variance_params}, we present the network-reported prediction uncertainties $\sigma_{[\rm Eu/H]}$ for the entire stellar sample, regardless of the quality cuts. These uncertainties are significantly higher and showing stronger systematics compared to Figs.~\ref{fig:val_variance_params} and~\ref{fig:best_practices_variance_params}, motivating the quality cuts and selection of the ``golden sample''.

Some outliers with very uncertain predictions ($\sigma_{\text{[Eu/H], pred}}>0.6$) are found in both the giant (1021 stars) and dwarfs (395 stars) samples. Those outliers occur mostly on the extreme edges of the temperature distribution: most (96\%) of the outlier giants have $T_{\rm eff} < 4000$ K, while most (64\%) of the outlier dwarfs have $T_{\rm eff} > 6000$ K (with a further 14\% having $T_{\rm eff} < 4000$ K). Furthermore, 97\% of the giant outliers trigger the {\tt continuum\_flux\_flag}. Dwarf outliers are caused by weak or entirely missing Eu lines: 99.7\% (all but one) do not pass the {\tt line\_depth} $> 0.02$ cut (with the remaining dwarf barely passing with {\tt line\_depth} $= 0.023$). Strikingly, 81\% of dwarf outliers have \textit{negative} {\tt line\_depth}.

After applying the recommended data practices (Section \ref{subsec:data_cuts}), the median $\sigma_{\text{[Eu/H], pred}}$ is 0.08 dex for giants and 0.10 dex for dwarfs. We note that applying the {\tt line\_depth} $> 0.02$ cut alone reduces the median $\sigma_{\text{[Eu/H], pred}}$ for the dwarf dataset from 0.16 to 0.10 dex.

These observations support the use of recommended data cuts; however, to allow the users maximum freedom and flexibility, we make predictions for the entire dataset of 61215 giants and 57731 dwarfs available.

\bsp	
\label{lastpage}
\end{document}